\documentclass[useAMS,usenatbib,usegraphicx]{mn2e}
 
\title[Evolution of Massive Galaxies]{The Properties and Evolution of a K-band Selected Sample of Massive Galaxies at $z \sim 0.4 - 2$ in the Palomar/DEEP2 Survey}
\author[Conselice et al.]{C. J. Conselice$^{1,2}$\thanks{E-mail:
conselice@nottingham.ac.uk}, K. Bundy$^{3,2}$, I. Trujillo$^{1}$, A. Coil$^{4,5}$, P. Eisenhardt$^{6}$, R.S. Ellis$^{7}$, \newauthor A. Georgakakis$^{8}$, J. Huang$^{9}$, J. Lotz$^{10}$, K. Nandra$^{8}$, J. Newman$^{11}$, \newauthor C. Papovich$^{4,12}$, B. Weiner$^{13,4}$, C. Willmer$^{4}$ \\
$^{1}$University of Nottingham, School of Physics \& Astronomy, Nottingham, NG7 2RD UK \\
 $^2$Previous address: Palomar Observatory, Caltech, MC 105-24 \\
 $^3$Department of Astronomy, University of Toronto, Canada \\
 $^4$Steward Observatory, University of Arizona, Tuscon, AZ \\
 $^5$Hubble Fellow \\
 $^6$Jet Proportion Laboratory, Caltech, Pasadena, CA \\
 $^7$California Institute of Technology, MC 105-24, Pasadena, CA 91125 \\
 $^8$Imperial College, London \\
 $^9$Harvard-Smithsonian Center for Astrophysics, Cambridge, MA \\
 $^{10}$NOAO, Tuscon, AZ \\
 $^{11}$Lawrence Berkeley National Laboratory, Berkeley CA \\
 $^{12}$Spitzer Fellow \\
 $^{13}$University of Maryland, College Park MD}

\def\solm{M$_{\odot}\,$}

\def\solm{M$_{\odot}\,$}

\def\mass{$10^{11}$ M$_{\odot}\,$}
\def\hmass{$10^{11.5}$ M$_{\odot}\,$}
\def\lmass{$10^{10.5}$ M$_{\odot}\,$}

\begin{document}

\date{Accepted ; Received ; in original form}

\pagerange{\pageref{firstpage}--\pageref{lastpage}} \pubyear{2002}

\maketitle

\label{firstpage}

\begin{abstract}

We present the results of a study on the  
properties and evolution of massive (M$_{*} > 10^{11}$ \solm) galaxies at
$z \sim 0.4 - 2$ utilising Keck spectroscopy, Near-Infrared 
Palomar imaging, and Hubble, Chandra, and Spitzer data covering
fields targeted by the DEEP2 galaxy spectroscopic survey.  
Our sample is K-band selected and stellar mass limited, based on wide-area 
near-infrared imaging from the Palomar Observatory Wide-Field Infrared
Survey, which covers
1.53 deg$^{2}$ to a 5 $\sigma$ depth of $K_{\rm s,vega} \sim 20.5$. 
Our primary goal is obtaining a broad census of massive galaxies
through measuring how their number and mass densities,
morphology, as well as their star formation and AGN content evolve
from $z \sim 0.4 - 2$.   Our major findings
include: (i) statistically  the mass and
number densities of M$_{*} >$ \mass galaxies show little evolution between
$z = 0 - 1$, and from $z \sim 0 - 2$ for M$_{*} >$ \hmass galaxies.  We however
find  significant evolution between $1 < z < 1.5$ for  
\mass $<$ M$_{*} <$ \hmass 
galaxies. (ii) After
examining the structures of our galaxies using Hubble ACS imaging, we find
that M$_{*} >$ \mass selected galaxies show a nearly constant elliptical 
fraction 
of $\sim 70-90\%$ at all redshifts. The remaining objects tend to be 
peculiars possibly undergoing  mergers at $z > 0.8$, while spirals dominate
the remainder at 
lower redshifts. A significant fraction ($\sim$ 25\%) of these early-types 
contain
minor structural anomalies. (iii) We find that only a fraction ($\sim 60$\%)
of massive galaxies with M$_{*} >$ \mass are on the red-sequence at 
$z \sim 1.4$, while nearly 100\% evolve onto it by $z \sim 0.4$. 
(iv) By utilising Spitzer MIPS imaging and [OII] line fluxes we argue that 
M$_{*} >$ \hmass galaxies have a steeply declining star formation rate 
density $\sim (1+z)^{6}$. By examining the contribution of star formation to
the evolution of the mass function, as well as the merger history through
the CAS parameters, we determine that M$_{*} >$ \mass galaxies undergo
on average $0.9^{+0.7}_{-0.5}$ major mergers at $0.4 < z < 1.4$.  
(v) We find that a high (5\%) fraction
of all M$_{*} >$ \mass galaxies are X-ray emitters.  Roughly half of these
are morphologically distorted ellipticals or peculiars. Finally, we
compare our mass growth with semi-analytical models from the Millennium
simulation, finding relative good agreement at $z < 2$ for the 
M$_{*} < 10^{11.5}$ \solm systems, but that the number and
mass densities of  M$_{*} > 10^{11.5}$ \solm galaxies are under predicted
by a factor of $> 100$.

\end{abstract}


\section{Introduction}

Understanding when and how the most massive galaxies in the universe
formed is one of the most outstanding problems in cosmology and galaxy 
formation.   Massive galaxies are predicted in Cold Dark Matter based 
models of structure formation to form gradually with time through the 
merging of smaller systems (e.g,. White \& Rees 1978).  While there is 
some evidence for this 
process (e.g., Le Fevre et al. 2000; 
Patton et al. 2002; Conselice et al. 2003a,b; Bridge et al. 2007), many
details are still lacking. Alternatively, massive galaxies, which
are mostly ellipticals in today's universe (e.g., Conselice 2006a), 
may have formed in a very rapid collapse of gas (e.g., Larson 1974).  
Observational evidence suggests that passively evolving massive galaxies 
exist at $z \sim 1$, as well as at even early times, $z > 2$ (e.g., Dunlop et 
al. 1996; Spinrad et al. 1997; Fontana et al. 2004; Daddi et al. 2004; 
Glazebrook et al. 2004; Treu et al. 2005; Labbe et al. 2005;  Papovich et 
al. 2006;  Grazian et al. 2006; Kriek et al. 2006; Lane et al. 2007). These 
systems are a subset of the massive galaxy population at high redshift, yet 
their nature, and whether or not they can account for all of the most massive 
galaxies in today's universe is still unknown.  We also do not yet know if 
these massive galaxies are still forming when the universe 
was half its current age (at $z \sim 1$). 

 Massive galaxies
are largely the test-bed for galaxy models, and therefore 
understanding their evolution observationally
is an important test of the physics behind galaxy formation.  A major
part of this problem is determining when massive galaxies  form.  Some
studies claim that massive galaxies are all in place by $z \sim 1$ (e.g.,
Glazebrook et al. 2004). However, since star 
formation and merging activity has been seen in ellipticals from $z \sim 0$ to
$z \sim 1$ (Stanford et al. 2004; Lin et al. 2004; Teplitz et al. 2006), a 
definite answer remains elusive.
If it were possible to date every star in nearby massive galaxies, we 
could in principle determine the 
formation epoch and time-scales of these systems by
examining their individual stars.  We cannot however resolve all the stars in 
galaxies, and their integrated stellar properties, such as colours, become 
degenerate after about 5 Gyrs (e.g., Worthey 1994).   Stellar
ages in massive galaxies also do not necessarily correlate with the
assembly of mass through, for example, merging activity (Conselice
2006b; De Lucia et al. 2006; Trujillo et al. 2006).  An alternative approach towards understanding 
massive galaxies and their evolution is empirically measuring the number 
densities, 
morphologies, star formation rates, and stellar masses of the most 
massive systems at some fiducial time and to compare these to similar
quantities at different times (redshifts), and with models.

This has however traditionally been a challenging problem
since the most massive galaxies are typically very red, either due to evolved
stellar populations, or by formation in dusty starbursts (e.g., Graham
\& Dey 1996).  Because of the shape of the spectral energy distributions of
evolved galaxies, they are identifiable at $z \sim 1$ due to 
their extremely red near-infrared to optical colours.  These objects
are sometimes known as extremely red objects (Elston et al. 1988), which
sample a fraction of $z > 1$ massive galaxies (e.g., Moustakas et al. 2004;
Conselice et al. 2007a). 
Conversely if one can obtain complete redshift samples, and combine
these with deep near-infrared imaging, it is possible to locate
the most massive galaxies up to $z \sim 1.4$ without relying on 
assumptions concerning their spectral energy distributions, morphologies, or
star formation histories.

Tracing the total evolution of stellar mass through cosmic time has been 
attempted in several fields, including the Hubble Deep Fields (Dickinson et al.
2003), the GEMS/COMBO-17 field (Borch et al. 2006), and the GOODS fields (Bundy
et al. 2005).  Recently there have also been
claims for the determination of the stellar mass evolution of galaxies
at $z \sim 0 - 3$ (Dickinson et al. 2003;  Fontana
et al. 2004; Bundy et al. 2006; Rudnick et al. 2006).  Some of these studies 
claim that many massive
galaxies are formed by $z \sim 2-3$, and that these systems prove to be
a complication for Cold Dark Matter models of structure formation 
(e.g., Glazebrook et al. 2003).  There are however
several limitations
to these studies. The two most important being the use of photometric
redshifts, and the small field sizes used in these earlier 
studies.  What is now
needed is a large spectroscopic survey over a large area, with deep 
near-infrared
imaging to measure accurate stellar masses of galaxies found when the universe
was younger than half its current age.  We provide such a study in this
paper by utilising $\sim$ 1.5 deg$^{2}$ Palomar NIR imaging of the
DEEP2 fields (Davis et al. 2003). We also go beyond 
previous work by utilising ancillary data on a purely mass selected sample
to determine the morphologies, colours, and star formation rates
of massive galaxies, and how these properties evolved down to $z \sim 0.4$.

Our galaxies are selected from the Palomar Observatory Wide-Field
Infrared survey, which is
designed to determine robustly the stellar mass evolution of galaxies, and
identify the most massive galaxies, at $0.4 < z < 2$.  Previously, we have
utilised spectroscopic  redshifts from the DEEP2
redshift survey (Davis et al. 2003; Faber et al. 2005) 
to probe the mass function down to M$^{*}$+3 (Bundy et al. 2005,
2006), and to study the properties of distant red galaxies (Conselice
et al. 2007).     Our goal in this paper is specifically
to understand the evolutionary history and properties of high
redshift massive galaxies, defined as systems with M$_{*} > 10^{11}$ \solm\,
at $z < 2$.   One of our primary objectives is to study systems at $z < 1.4$,
where we have ancillary information from other telescopes. 

This paper is organised as follows.  \S 2 is a presentation of
our data.  \S 3 gives an overview of the methodology
used to determine the quantities used in this paper, while \S 4 is
a discussion of the properties and evolution of massive galaxies at 
$0.4 < z < 1.4$. \S 5 is a discussion of our results in terms of
galaxy evolution scenarios, and \S 6 is a summary.
Throughout this paper we use a standard cosmology of H$_{0} = 70$ km s$^{-1}$
Mpc$^{-1}$, and $\Omega_{\rm m} = 1 - \Omega_{\lambda} = 0.3$. All quoted
magnitudes are in the Vega systems, unless otherwise specified.

\section{Data}

The galaxies we study in this paper consists of those in the fields covered by
the Palomar Observatory Wide-Field Infrared Survey (POWIR, Table~1), 
excluding the GOODS North field discussed in Bundy et al. (2005).  The 
POWIR survey was designed to obtain deep K-band and J-band data over a 
significant ($\sim$1.5 deg$^2$) area.   Observations were 
carried out between September 2002 and October 2005 over a total of $\sim
70$ nights. This survey
covers the GOODS field North (Giavalisco et al. 2004; Bundy et al. 2005),
the Extended Groth Strip (Davis et al. 2006), and three other fields 
that the DEEP2 team
has observed with the DEIMOS spectrograph (Davis et al. 2003).  The total
area we cover in the K-band is 5524 arcmin$^{2}$ = 1.53 deg$^{2}$, with
half of this area imaged in the J-band. Our goal depth was K$_{\rm s,vega} =
21$, although not all fields are covered this deep, but all have 5$\sigma$
depths between K$_{\rm s,vega} = 20.2 - 21.5$ for point sources, measured in a
2\arcsec\, diameter aperture.  Table~1 lists the DEEP2 fields, and the area 
we have imaged in each.  For our purposes we
will abbreviated the fields covered as: EGS (Extended Groth Strip),
Field 2, Field 3, and Field 4.
 
\setcounter{table}{0}
\begin{table}
 \caption{The Palomar Fields and WIRC pointings Areas}
 \label{tab1}
 \begin{tabular}{@{}lccc}
  \hline
Field & RA & Dec. & area (arcmin$^{2}$) \\
\hline
EGS & 14 17 00 & +52 30 00 & 2165 \\
Field 2 & 16 52 00 & +34 55 00 & 787 \\
Field 3 & 23 30 00 & +00 00 00 & 984 \\
Field 4 & 02 30 00 & +00 00 00 & 984 \\
\hline
 \end{tabular}
\end{table}

All of our K$_{\rm s}$-band data were acquired utilising the WIRC camera on the Palomar
5 meter telescope.  WIRC has an effective field of view of 
$8.1\arcmin \times 8.1\arcmin$, with a pixel scale
of 0.25\arcsec pixel$^{-1}$.  Our total survey contains 75 WIRC pointings.
During the K$_{\rm s}$-band observations we used 30 second integrations,
with four exposures per pointing.  The J-band observations were taken
with 120 second exposures per pointing. Typical total exposure times were 
between one and two hours for both bands. Our reduction procedure follows
standard method for combining NIR ground-based imaging, and is described
in more detail in Bundy et al. (2006). The resulting seeing FWHM in the 
K$_{\rm s}$-band imaging ranges from 0.8'' to 1.2'', and is typically 1.0''.  

Photometric calibration was carried out by referencing Persson et al. 
(1998) standard stars during photometric conditions.  
The final images were made by combining individual mosaics obtained over
several nights. The K$_{\rm s}$-band mosaics are comprised of coadditions of 
$4 \times 30$ 
second exposures dithered over a non-repeating 7.0'' pattern.  The images
were processed using a double-pass reduction pipeline we developed
specifically for WIRC. For galaxy detection and photometry 
we utilised the SExtractor package (Bertin \& Arnouts 1996). False
artifacts are removed through SExtractor flags which identify sources
that do not have normal galaxy or stellar profiles.
From this we construct a K-selected sample, which is then cross-referenced 
with the DEEP2 redshift catalog.

Other data used in this paper consists of: optical
imaging from the CFHT over all fields, MIPS imaging from
the {\em Spitzer Space Telescope}, imaging from the Advanced Camera for 
Surveys (ACS) on Hubble, Chandra X-ray imaging, and spectroscopy from the 
DEIMOS
spectrograph on the Keck II telescope (Davis et al. 2003).  A summary
of these ancillary data sets, which are mostly within the Extended Groth Strip,
are presented in Davis et al. (2006).

The optical imaging of our fields comes from the CFHT 3.6-m, and consists
of data in the B, R and I bands taken with the CFH12K camera - a 12,288 
$\times$ 8,192 pixel CCD mosaic with a pixel scale of 0.21\arcsec.  
The integration times for these observations are 1 hour in $B$ and $R$, and
2 hours in $I$, per pointing, with a R-band 5 $\sigma$ depth of 
$R_{\rm AB} \sim 25.1$, and similar depths at $B$ and $I$ (Coil et al. 
2004b).    The details of the data reduction
for this data is described in Coil et al. (2004b).  From this imaging
data a R$_{\rm AB}$ = 24.1 
magnitude limit was used for determining targets for the DEEP2 spectroscopy.  
The details for how these imaging data were acquired and reduced are
covered in Coil et al. (2004b).  The seeing for the optical imaging is
roughly the same as that for the NIR imaging, and we measure photometry
consistently using a 2\arcsec\, diameter aperture.
  
The Keck spectra were acquired with the DEIMOS spectrograph 
(Faber et al. 2003) as part of the DEEP2 redshift survey (Davis
et al. 2003).  The selection for targets for the DEEP2 spectroscopy
was based on the optical properties of the galaxies detected in the
CFHT photometry, with the basic selection criteria $R_{\rm AB} < 24.1$.   
Spectroscopy in the EGS was acquired through
this magnitude limit, with no strong colour cuts applied to the selection.  
Objects in Fields 2-4 were selected for spectroscopy based on their position in
$(B-R)$ vs. $(R-I)$ colour space to focus on galaxies at redshifts $z > 0.7$.  
The total survey includes over 30,000 galaxies with a secure redshift, with 
about a third of these in the EGS field.  In all fields
the sampling rate for galaxies that meet the selection criteria is
60\%.

The DEIMOS spectroscopy was obtained using the
1200 line/mm grating, with a resolution R $\sim 5000$ covering
the wavelength range 6500 - 9100 \AA.  Redshifts were measured through
an automatic method comparing templates to data, and we only utilise
those redshifts measured when two or more lines were
identified, providing very secure measurements.  Roughly 70\% of all
targeted objects result in secure redshifts. Most of the redshift
failures are galaxies at higher redshift, $z > 1.5$ (Steidel et al.
2004), where the [OII] $\lambda$3727 line leaves the optical window. 

The ACS imaging over the EGS field covers a 10.1\arcmin $\times$ 70.5\arcmin\,
strip, for a coverage area of 0.2 deg$^{2}$.  This ACS imaging
is discussed in Lotz et al. (2006), and is briefly described
here.  The imaging consists of 63 tiles imaged in both the
F606W (V) and F814W (I) bands.  The 5-$\sigma$ depths reached in these
images are V = 26.23 (AB) and I = 27.52 (AB) for point
sources, and about two magnitudes brighter for extended objects.

We further utilise Spitzer MIPS 24$\mu$m imaging, and Chandra X-ray imaging
of the EGS field (Nandra et al. 2007) to determine
the star forming and AGN properties of our sample.  The MIPS 24$\mu$m imaging
is part of the IRAC team GTO program (Papovich et al. 2004).  Our procedure
was to use a catalog of 24$\mu$m sources matched to our list of
massive galaxies within 1\arcsec\,, and brighter than 
$f_{24\mu m} = $60$~\mu$Jy.  
This ensures that the source matching is reliable, and that the MIPS detections
are significant to $> 3$ $\sigma$ confidence.  We then convert these
24$\mu$m fluxes into total infrared fluxes utilising the fact that there
is a good correlation between fluxes at 15$\mu$m and the total IR flux
(Le Floc'h et al. 2005).  We use various templates to convert our
observed 24$\mu$m flux into a total IR flux, from which we compute
star formation rates (\S 4.4.2).

Our matching procedures for these catalogs progressed in the manner
described in Bundy et al. (2006) with the K-band catalog serving as our
reference. We then match the optical catalogs and spectroscopic
catalogs to the K-band catalog, after correcting for astrometry by
referencing all systems to 2MASS stars.  The MIPS catalogs and
X-rays catalogs are likewise matched in a similar manner.

\subsection{Photometric Errors and Detection}

We  estimate photometric errors, and the K-band detection limit of each 
K-band image by randomly inserting fake objects of known magnitude, surface
brightness profile, and size into 
each image, and then recovering these simulated objects with the same 
detection parameters
used for real objects.   The most basic
simulations we perform are done by simply inserting  objects with 
Gaussian profiles with a FWHM of 1\farcs3 to approximate the shape of 
slightly extended, distant galaxies. We then retrieve these objects using the
same SExtractor parameters used for the original galaxy detections. We use
these simulated retrievals to determine the upper limit
completeness of our sample, as well as constrain errors on our 
photometric measurements.   Actual galaxy profiles however have
extended envelopes that could be more difficult to measure accurately than
compact Gaussian profiles (e.g., Graham et al. 2005).

To determine the detection and photometric fidelity of our images in more 
detail, we create two sets of 10,000 mock galaxies, each with an intrinsic 
exponential and de Vaucouleurs profile. These simulated galaxies
have properties which are uniformly distributed as 
follows: K$_{\rm s}$ band total magnitudes between 15.5
and 20.5 mag, effective radius R$_{\rm e}$ between 0.0625 and 3.75 
arcsec (equivalent to 0.5-30 kpc at z$\sim$1), and ellipticities  
between 0 and 0.8.  These simulated sources are placed randomly on our
Palomar images, and extracted in the same manner as 
the real source detections. We construct from these simulations
detection maps that reveal the fraction of input 
artificial sources detected per input magnitude and input log (re) 
bin (see Figure~1). 
As expected, galaxies with a de Vaucouleurs profile, which are 
more centrally concentrated, are easier to detect at a given magnitude.
Note that the detection fractions plotted in Figure~1 are independent 
of output magnitude and size, and are plotted based on their
input parameters.

We also estimated systematic errors in measuring magnitudes
due to our detection method by using the same simulations (see Figure~2). 
To do this, we compute the magnitude difference between the recovered  
and  input sources within the input magnitude and size bin. To estimate 
the output magnitudes, we use exactly the same methods used to find and 
measure photometry of our actual 
galaxies. The decreased surface brightness of objects, at a given input  
magnitude, results in output magnitudes systematically fainter at 
larger effective radii. In particular, due to the larger tail of the de 
Vaucouleurs profiles, at a given magnitude and size, the difference in 
magnitude is larger for the de Vaucouleurs profiles than for the 
exponential models.  At the range of sizes for our sample objects, which
range from 0.3''-0.7'' (Trujillo et al. 2007, submitted), we find that the 
magnitude and recovery fraction are
essential 100\% of their simulated value. Depending on the sizes of our 
galaxies, we could
be missing as much as 0.2 mag in the recovered K-band light. We will
address this in more detail in \S 4.2.1.

\section{Methods}

We utilise several methods to study the properties of our K-band
selected,
stellar mass limited sample.  The outline of the procedure
is described below.  The first step in this 
process, after reducing the K-band imaging,
is to create K-selected catalogs. We make these catalogs using
SExtractor, optimised for detection and splitting, with both topics
discussed in Bundy et al. (2006, 2007).   We then utilise the 
K-band imaging, combined with the optical imaging, to compute stellar masses
from which our sample is selected.  We then match these massive galaxies
to MIPS, X-ray and Hubble sources from which we derive the physical
features, and the evolution of these systems.

\subsection{K-band Selection}

Within the total K-band area of our survey (1.53 deg$^{2}$) we 
detect 61,489 sources, 
after removing false artifacts.  Most of these objects (92\%) are at K $< 21$,
while 68\% are at K $< 20$, and 37\% are at K $< 19$. In total there
are 38,613 objects fainter than K $= 19$ in our sample.  
Out of our total K-band population, 10,693 objects have 
secure spectroscopic redshifts from the DEEP2 redshift survey (Davis et 
al. 2003).  We supplement these by 37,644 photometric redshifts within
the range $0 < z < 2$ (\S 3.2).

Because our spectroscopy is R-band selected, while our stellar masses
are based on K-band detections, we are required to divided our
sample into different sub-samples.  We 
construct a ``primary sample'' consisting of those galaxies
with both spectroscopic redshifts and K-band detections. By
construction, these galaxies also have optical magnitudes  $R_{\rm AB} < 24.1$.
The matching between our K-band catalogue, and the DEEP2 catalogue
was done with a 1\arcsec\, tolerance. This results in a very low
spurious rate of 1-2\% due to the low surface densities in both
catalogues.  Most of the galaxies with spectroscopic redshifts
are detected in the K-band, at a rate between 65-90\%, depending on the K-band
depth.  Because we are interested in the most massive galaxies, which appear
bright in the NIR, the present study is not biased by these non-matches.  

The ``secondary'' sample consists of those galaxies with photometric
redshifts. Within the secondary sample, there are two kinds of 
photometric redshifts. The first are those systems which have 
optical magnitudes at $R_{\rm AB} < 24.1$, the limit for spectroscopy.
The other types of photometric redshifts are for those galaxies which have
optical magnitudes fainter than this optical limit, in which we use
a different method to compute photometric redshifts (\S 3.2).  This
second set includes all galaxies at $z > 1.4$, which
have purely photometrically measured redshifts.

\subsection{Photometric Redshifts}

We calculate photometric redshifts for our K-selected galaxies which do
not have DEEP2 spectroscopy.  This sample is 
referred to as the photometric-redshift or secondary sample.  
These photometric redshifts are based on
the optical+near infrared imaging, in the BRIJK (or BRIK for half the
sample) bands, and are fit in two ways, 
depending on the brightness of a galaxy in the optical.   For galaxies that
meet the spectroscopic criteria, $R_{\rm AB} < 24.1$, we utilise a neural
network photometric redshift technique to take advantage of the
vast number of secure redshifts with similar photometric data.  Most
of the $R_{\rm AB} < 24.1$ sources not targeted for spectroscopy should be 
within our redshift range of interest, at $z < 1.4$.    The neural network 
fitting is 
done through 
the use of the ANNz (Collister \& Lahav 2004) method and code.
To train the code, we use the $\sim 5000$ redshifts in the EGS, which
has galaxies spanning our entire redshift range.  The training of the 
photometric
redshift fitting was in fact only done using the EGS field, whose
galaxies are nearly completed selected based on the magnitude
limit of $R_{\rm AB} < 24.1$. We then use this training to calculate the
photometric redshifts for galaxies with $R_{\rm AB} < 24.1$ in all fields.   
The overall agreement between our photometric redshifts and our ANNz 
spectroscopic redshifts is very good 
using this technique, with $\delta z/(1+z) = 0.07$ out
to $z \sim 1.4$. The agreement is even better for the M$_{*} >$ \mass
galaxies where we find $\delta z/(1+z) = 0.025$ across all of our
four fields.   The photometry we use for our photometric
redshift measurements are done with a 2\arcsec\, diameter aperture.

For galaxies which are fainter than $R_{\rm AB} = 24.1$ we utilise photometric
redshifts using Bayesian techniques, and the software from
Benitez (2000).  For an object to have a photometric redshift
we require that it be detected at the 3-$\sigma$ level in all
optical and near-infrared (BRIJK) bands, which in the R-band
reaches $R_{\rm AB} \sim 25.1$ (Coil et al 2004b). We 
optimised our results, and corrected for systematics, through 
the comparison with spectroscopic redshifts, resulting
in a redshift accuracy of $\delta z/z = 0.17$ for $R_{\rm AB} > 24.1$ systems. 
These $R_{\rm AB} > 24.1$ galaxies are however only a very small part of our
sample. Up to $z \sim 1.4$ only 6 (2.6\%) of our M$_{*} >$ \hmass galaxies 
are in this regime, while 412 \mass $<$ M$_{*} <$ \hmass (9\%) of our
galaxies have an $R$-band magnitude this faint. All of these systems are 
furthermore at $z > 1$.  At $z > 1.4$ all of our sample galaxies 
are measured through the Benitez (2000) method due to a lack of
training redshifts.
There are instances where not all galaxies are detected in the
optical bands, and the way we deal with these galaxies is
discussed in \S 4.2.1 in terms of our measured number and mass densities.

One concern with utilising the DEEP2 spectroscopy as the prior
in the ANNz method is that any galaxies which are actually at
higher redshifts, but with $R_{\rm AB} < 24.1$, could contain a very incorrect
redshift. This is particularly important in this paper, as we
do not wish to have additional galaxies added to our sample
which could produce false artifacts in the number and mass density
evolution, as well as incorrect identifications of massive
galaxies.  We know that about 30\% of galaxies with $R_{\rm AB} < 24.1$ will
be at $z > 1.4$ (Steidel et al. 2004). These systems at $z > 1.4$
are however nearly all
fainter than $K = 20$ (Reddy et al. 2006), and are thus not likely  
contaminating our massive galaxy sample, which is nearly all at $K < 19$.

After our photometric redshifts were calculated,
the CFHT legacy survey published their own independent photometric
redshifts in the EGS, which are on average within $\delta z/(1+z)$ = 0.2
of our photometric redshifts.  The poorer agreement between the
CFHT legacy survey photometric redshifts and ours is likely due
to the lack of NIR data in the CFHT fits, and the more restrictive 
priors used in the CFHT photometric
redshift fitting (cf. Bundy et al. 2006).  In any case, we do not use 
individual photometric
redshifts, and only utilise them in large ensembles, where errors due
to photometric redshift mismatches can be properly accounted for.

\subsection{Stellar Masses}

We match our K-band selected catalogs to the CFHT optical data to obtain
spectral energy distributions (SEDs) for all of our sources, resulting in
measured BRIJK magnitudes.   From these we compute stellar masses based on 
the methods and results
outlined in Bundy, Ellis, Conselice (2005) and Bundy et al. (2006). Stellar
masses computed in this way have a long established 
history (e.g., Sawicki \& Yee 1998). About 40\% of our sample do not contain
J-band imaging, but the stellar masses computed do not vary significantly
when the J-band data is included or excluded.   All our stellar masses
are furthermore normalised by the observed rest-frame K-band light, which
is roughly at rest-frame $\sim 1 \mu m$ for most galaxies.

The basic mass fitting method consists of fitting a grid of model SEDs 
constructed
from Bruzual \& Charlot (2003) (BC03) stellar population synthesis models, with
different star formation histories. We use an exponentially declining model
to characterise the star formation history, with various ages, 
metallicities and dust contents included.  These models are parameterised
by an age, and an e-folding time for parameterising the star formation 
history, where SFR $\alpha\, e^{\frac{t}{\tau}}$.  The values of $\tau$ are
randomly selected from a range between 0.01 and 10 Gyr, while the age
of the onset of star formation ranges from 0 to 10 Gyr. The metallicity
ranges from 0.0001 to 0.05 (BC03), and the dust content is parametrised
by $\tau_{\rm V}$, the effective V-band optical depth for which we use values
$\tau_{\rm V} = 0.0, 0.5, 1, 2$.     Although we vary several parameters,
the resulting stellar masses from our fits do not depend strongly on the
various selection criteria used to characterise the age and the metallicity
of the stellar population.

It is however important to realise that these  parameterisations are
fairly simple, and it remains possible that stellar mass from
older stars is missed under brighter, younger, populations. While
the majority of our systems are passively evolving older stellar populations,
it is possible that up to a factor of two in stellar mass is missed in any 
star bursting blue systems.  However, stellar masses measured through
our technique are roughly the expected factor of 5-10 smaller than
dynamical masses at $z \sim 1$ using a sample of disk galaxies
(Conselice et al. 2005b), demonstrating their inherent reliability.

We match magnitudes derived from these model 
star formation histories to the actual data to obtain a measurement
of stellar mass using a Bayesian approach.
We calculate 
the likely stellar mass, age, and absolute magnitudes for each galaxy at all 
star formation histories, and determine stellar masses based on this
distribution.  Distributions with larger ranges of stellar masses
have larger resulting uncertainties. It turns out that while parameters such 
as the age, e-folding time, metallicity, etc. are not likely accurately fit
through these
calculations due to various degeneracies, the stellar mass is robust.  
Typical errors for our stellar masses are 0.2 dex from the width
of the probability distributions.  There are also uncertainties from
the choice of the IMF.  Our stellar masses utilise the
Chabrier (2003) IMF, which can be converted to Salpeter IMF stellar masses 
by adding 0.25 dex.  There are additional
random uncertainties due to photometric errors.  The resulting
stellar masses thus have a total random error of 0.2-0.3 dex,
roughly a factor of two.  The details behind these mass measurements
and their uncertainties is also described in papers such
as Brinchmann \& Ellis (2000), Papovich et al. (2006) and
Bundy et al. (2006).

There is furthermore the issue of whether or not our stellar masses
are overestimated based on using the Bruzual \& Charlot (2003)
models.  It has recently been argued by Maraston (2005) and
Bruzual (2007) that a refined treatment of thermal-pulsating AGB stars 
in the BC03 models results in calculated stellar masses that can be
too high by a factor of a few. While we consider an uncertainty
of a factor of two in our stellar masses, it is worth investigating
whether or not our sample is in the regime where the effects of
a different treatement of TP-AGB stars in e.g., Maraston (2007) will 
influence our mass measurements.  This
has been investigated recently in Maraston (2005) and Bruzual
(2007) who have both concluded that galaxy stellar masses computed
with an improved treatment of TP-AGB stars are roughly 
50-60\% lower.

This problem has also been recently investigated independently
by Kannappan \& Gawiser (2007) who come to similar conclusions,
but do not advocate one model over another.  Furthermore, the effect
of TP-AGB stars is less important at our rest-frame wavelengths probed 
than at longer wavelengths, especially in the rest-frame IR. Our survey 
is K-selected, and the
observed K-band is used as the flux in which the masses are computed. The
rest-frame wavelength probed with the observed K-band ranges from 0.7$\mu$m
to 1.5$\mu$m where the effects of TP-AGB stars are minimised.
The ages of our galaxies are also older than the ages where
TP-AGB stars have their most effect (Maraston 2005; Bruzual
2007).  To test this, after our analysis was finished, we utilised
the newer Bruzual and Charlot (2007, in prep) models, which include a new 
TP-AGB star prescription, on our massive galaxy sample. From this we
find on average a $\sim$ 0.07 dex smaller stellar mass using the newer
models.  At most, the influence of TP-AGB stars will decrease our
stellar masses by 20\%.  The effect of this would decrease
the number of galaxies within our sample, particularly those
close to the M$_{*} =$ \mass boundary.  This systematic error is however
much smaller than both the stellar mass error we assume (0.3 dex), and
the cosmic variance uncertainties, and thus we conclude it is not
a significant factor.

\section{The Properties and Evolution of Massive Galaxies}

\subsection{Sample Selection}

Our K-band selection allows us to measure the stellar masses of
galaxies to $K \sim 20-21$.  The stellar mass function for galaxies
within these limits is discussed in Bundy et al. (2006) for our sample.
In this paper we discuss only those galaxies with M$_{*} >$ \mass.   
The K-band magnitude distribution with redshift and the
stellar mass distribution with redshift are shown in Figure~3.
Figure~3 displays our total sample - consisting of the primary 
spectroscopic, and secondary photometric redshift selected
galaxies combined.   While the selection limit of M$_{*} =$ \mass is somewhat
arbitrary, it is however the mass limit in which we are reasonably
completed out to $z \sim 2$. 

Figure~3a shows the K-band magnitude
distribution for our sample, demonstrating that all of our sample
is at $K < 20$, where we are 100\% complete in K-band detections
(\S 2.1; Conselice et al. 2007b).  However, we are not complete in
other bands at all redshifts for a $K < 20$, and thus a M$_{*} >$ \mass 
selected sample. At $R_{\rm AB} = 25.1$ we are complete 
at \lmass $<$ M$_{*} <$ \mass
up to $z \sim 1$, up to $z \sim 1.4$ for \mass $<$ M$_{*} <$ \hmass,
and up to $z \sim 2$ for M$_{*} >$ \hmass.  However, this completeness
is not the same in the B-band.  This needs to
be accounted for as we do not include photometric redshifts measurements
for galaxies which are undetected in one of our bands (\S 3.2).
However, for the redshift and mass
ranges plotted in Figure~3 \& 4 and listed in Table~2, we are 100\% complete
at $z < 1.2$ and are 2\% - 10\% incomplete at higher redshifts. The
method we use to calculate, and correct for, this incompleteness in the 
number and mass densities is discussed in \S 4.2.1.

\begin{figure*}
 \vbox to 120mm{
\includegraphics[angle=0, width=184mm]{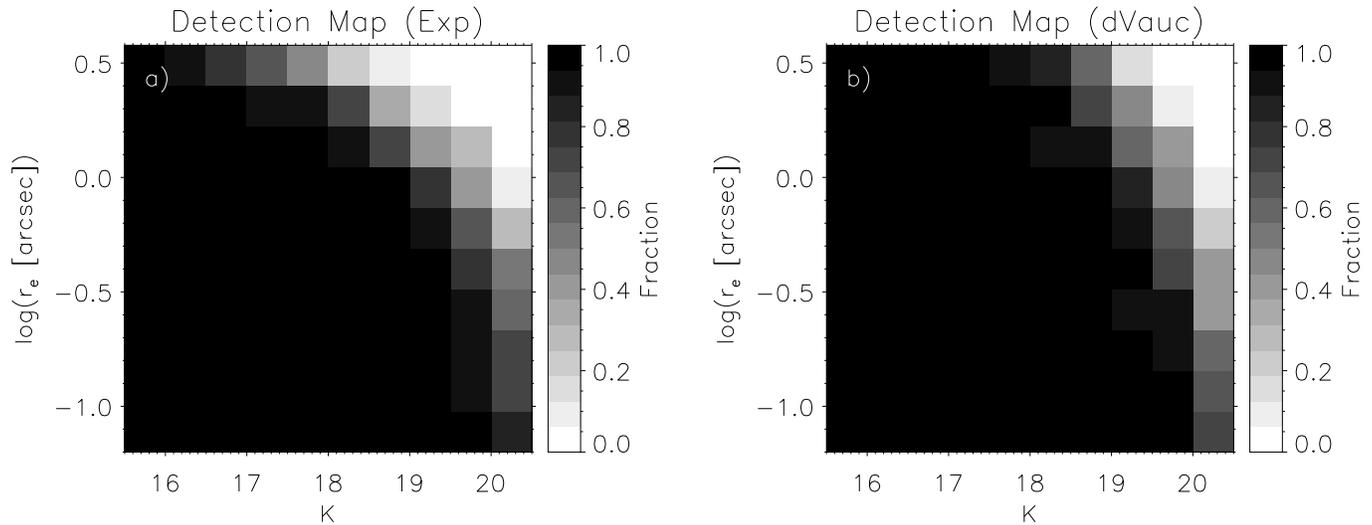}
 \caption{a) Detection map for simulated sources with exponential profiles 
placed at random in our Ks band images. The gray-scale 
map reflects
the ratio between input and recovered objects per input magnitude and log (re)
bin. b) Same as a) but showing the detection map for simulated sources with a
de Vaucouleurs profiles. }
} \label{sample-figure}
\vspace{-3cm}
\end{figure*}

Utilising our stellar mass catalogs, the selection of massive 
galaxies out to $z \sim 2$ is straightforward, and is simply a cut at 
stellar masses M$_{*} >$ \hmass and M$_{*} >$ \mass. We do not consider 
lower mass galaxies 
in this analysis, due to significant incompleteness, although these 
galaxies may show the most evolution within
these redshifts, as can be seen in the star formation
downsizing (e.g., Bundy et al. 2006).   In this paper,
galaxies with masses M$_{*}$ $<$ \mass are only discussed in terms of
their relationship to higher mass galaxies.  Note that our
massive galaxies are much brighter than the survey 5 $\sigma$ limit of 
$K = 20.5-21$.  As Figure~3 shows, we are unlikely missing galaxies within
our mass selection which are fainter than this limit up 
to $z \sim 2$ (\S 4.2.1).    For example, at 
$z \sim 1.5-2$ galaxies with a maximum M/L ratio with M$_{*} =$ \mass would 
still be several times brighter than our K-limit.

\setcounter{table}{1}
\begin{table*}
 \caption{Galaxy Number and Mass Density Evolution and Star Formation Rates as a Function of Mass}
 \label{tab1}
 \begin{tabular}{@{}ccccc}
  \hline
Mass & Redshift & log ($\phi$) (h$^{3}_{70}$ Mpc$^{-3}$ dex$^{-1}$)& log ($\rho$) (h$^{3}_{70}$ \solm Mpc$^{-3}$) & log $\rho_{\rm SFR}$ (\solm h$_{70}^{3}$ yr$^{-1}$ Mpc$^{-3}$) \\
\hline
     log M$_{*} > 11.5$   &      0.5 & -4.62$^{+0.51}_{-0.26}$  &    6.83$^{+0.49}_{-0.24}$  & -4.9$^{+0.2}_{-0.3}$ \\
   &      0.7    &   -4.29$^{+0.32}_{-0.21}$  &      7.17$^{+0.28}_{-0.21}$ & -3.6$^{+0.2}_{-0.3}$ \\
   &      0.9    &   -4.08$^{+0.23}_{-0.20}$  &      7.34$^{+0.28}_{-0.20}$ & -3.0$^{+0.2}_{-0.3}$ \\
   &      1.1   &    -4.28$^{+0.30}_{-0.20}$  &      7.10$^{+0.29}_{-0.20}$ & -2.9$^{+0.2}_{-0.3}$ \\
   &      1.3    &   -5.05$^{+0.45}_{-0.22}$  &      6.73$^{+0.24}_{-0.21}$ & -3.3$^{+0.2}_{-0.3}$ \\
   &      1.5     &  -4.91$^{+0.20}_{-0.22}$  &      6.83$^{+0.14}_{-0.21}$ & ... \\
   &      1.7    &   -5.03$^{+0.22}_{-0.22}$  &      6.66$^{+0.17}_{-0.22}$ & ... \\
   &      1.9     &  -5.33$^{+0.25}_{-0.25}$  &      6.36$^{+0.16}_{-0.25}$ & ... \\
\\
$11 <$ log M$_{*} < 11.5$ &  0.5    &  -3.32$^{+0.10}_{-0.11}$    &    7.76$^{+0.11}_{-0.11}$ & -2.6$^{+0.2}_{-0.3}$ \\
    &     0.7    &   -3.21$^{+0.09}_{-0.10}$  &       7.79$^{+0.10}_{-0.11}$ &  -2.2$^{+0.2}_{-0.3}$ \\
    &     0.9    &   -3.08$^{+0.08}_{-0.10}$  &       7.92$^{+0.09}_{-0.10}$ &  -2.0$^{+0.2}_{-0.3}$ \\
     &    1.1    &   -3.26$^{+0.09}_{-0.10}$  &       7.83$^{+0.09}_{-0.10}$ &  -1.9$^{+0.2}_{-0.3}$ \\
     &    1.3   &    -3.66$^{+0.09}_{-0.10}$  &       7.44$^{+0.08}_{-0.11}$ &  -2.1$^{+0.2}_{-0.3}$ \\
   &      1.5    & -3.80$^{+0.09}_{-0.11}$    &   7.42$^{+0.10}_{-0.11}$ & ... \\
   &      1.7    &  -4.10$^{+0.11}_{-0.11}$   &    7.11$^{+0.11}_{-0.11}$ & ... \\
\\
$10.5 <$ log M$_{*} < 11$    &    0.5       &   -2.80$^{+0.06}_{-0.07}$  &     7.88$^{+0.06}_{-0.07}$ & ... \\
    &     0.7     &     -2.80$^{+0.06}_{-0.07}$  &      7.88$^{+0.06}_{-0.07}$ & ... \\
    &     0.9     &     -2.69$^{+0.12}_{-0.07}$  &      7.79$^{+0.06}_{-0.07}$ & ... \\
\hline
 \end{tabular}
\end{table*}

\subsection{Number Densities of Massive Galaxies}

In this section we examine in detail the number density and
mass density evolution for massive galaxies found between
$z \sim 0$ and $z \sim 2$.
For the purposes of this paper we consider galaxies with 
M$_{*} > 10^{11}$ \solm as massive systems.  These galaxies are 
in fact the most massive galaxies in the nearby universe, and are
generally not found in smaller area NIR surveys.  For
example, in the Hubble Deep Field-North (HDF-N), there are only four galaxies 
with stellar masses M$_{*} >$ \mass, and none with M$_{*} >$ \hmass 
(Dickinson et al. 2003) using a 
Chabrier IMF.   In comparison, we have 4571 galaxies with $M_{*} >$ \mass,
and 225 galaxies with $M_{*} >$\hmass up to $z \sim 1.4$. Although the HDF-N 
might be slightly depleted of
massive galaxies, it shows the importance of using a large area
survey to determine the properties and evolution of massive
galaxies.   Roughly 30\% of our M$_{*}$ $>$\mass
galaxies have measured spectroscopic redshifts, while 36\% of the 
M$_{*}$ $>$\hmass galaxies 
have spectroscopic redshifts.  We show the redshift distribution of
our sample of galaxies as a function of redshift up to $z \sim 2$ as
a function of mass
and K-band magnitude in Figure~3, with the massive galaxies with masses
\mass $<$ M$_{*} <$ \hmass, and M$_{*} >$ \hmass labelled.

\begin{figure*}
 \vbox to 120mm{
\includegraphics[angle=0, width=184mm]{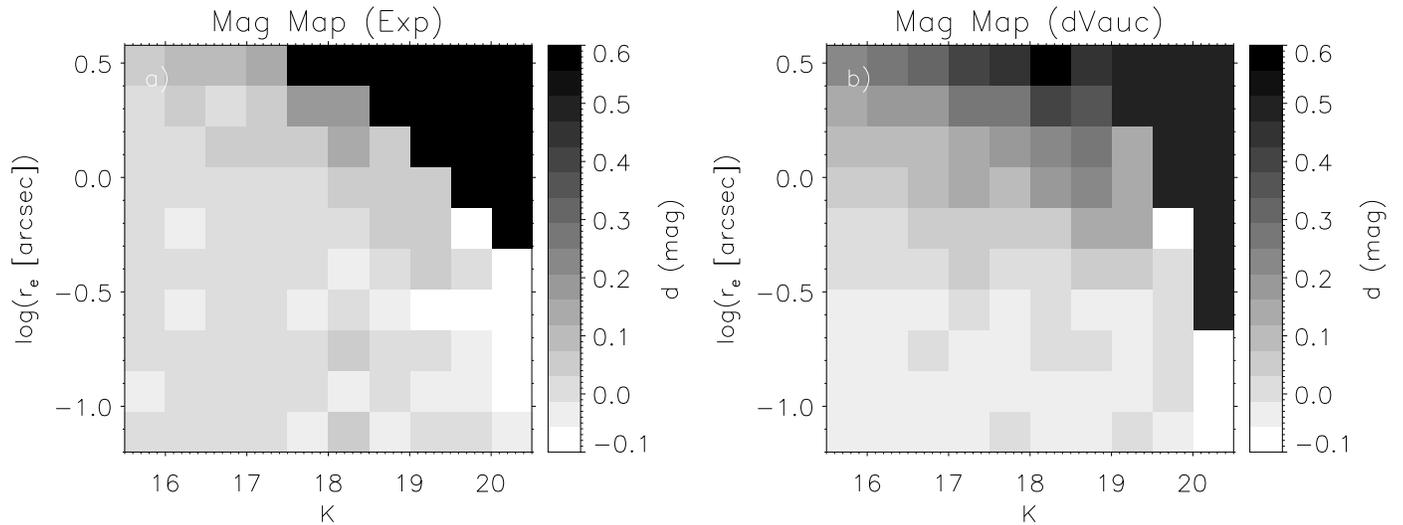}
 \caption{a) Magnitude map for simulated sources with exponential profiles 
placed at random in our Ks band images. The gray-scale 
map reflects
the difference between input and recovered magnitudes per input magnitude and
log (re) bin. b) Same as a) but showing the magnitude map for simulated sources
with a de Vaucouleurs profiles. }
} \label{sample-figure}
\vspace{-3cm}
\end{figure*}

The fact that such massive galaxies exist
at high redshift is not completely surprising, as a large number of
massive galaxies 
have been found at $z \sim 1.5-2$, albeit 
in much smaller fields (Glazebrook et al. 2004; Saracco et al. 2005).  Cosmic
variance is an issue in these previous studies, which have areas over a 
factor of ten smaller than ours. This is especially a problem, even within 
our survey, for the most massive galaxies, which are the most clustered within
the redshift ranges we examine (e.g., Coil et al. 2004a; 
Foucaud et al. 2007).    As a result, these 
massive M$_{*} >$ \hmass galaxies are nearly completely absent in previous
high redshift stellar mass studies, although a few examples exist in
some previous work using smaller area fields.

\subsubsection{Sources of Uncertainty and Redshift Completeness}

Because any variations in number or mass densities suggest evolution, it is
important to consider the various errors that can mimic real evolution in
our analysis both in terms of evolution in number/mass densities, as well as
for the selection of the most massive galaxy sample which we examine later
in the paper.
Beyond random and shot noise errors, we also consider stellar mass 
measurement errors, both systematic and random, as well as errors from cosmic 
variance when examining the evolution of galaxy number and mass densities.
This includes considering in the stellar mass errors issues with
photometry, both random and systematic.
We also consider in our stellar mass density and number density measurements
the errors produced through using photometric redshifts. We do this through
Monte-Carlo simulations of how the stellar mass function would change
given the known uncertainty on these measurements.  This also gives us
some idea of the Eddington bias which is affecting our selection near
the M$_{*} = $ \mass and M$_{*} = $ \hmass boundary selection limits.

\begin{figure*}
 \vbox to 120mm{
\includegraphics[angle=0, width=184mm]{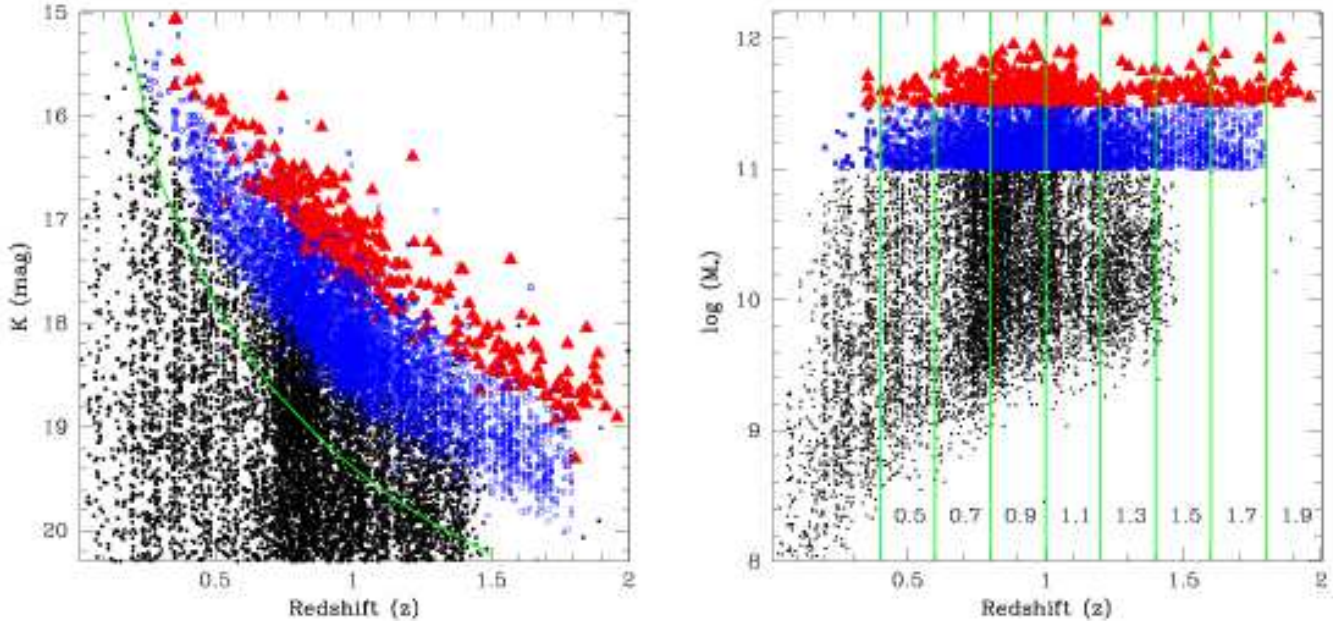}
 \caption{Left panel: the K magnitude vs. redshift ($z$) diagram
for our total $z < 2$ sample  with spectroscopic and photometric redshifts
combined.  
The red triangles show the locations of galaxies with derived stellar
masses M$_{*} >$ \hmass, and the blue boxes show the location of galaxies with 
\mass $<$ M$_{*} <$ \hmass. The 
solid green line shows the evolution of L$^{*}$ in the K-band
out to $z \sim 1.5$ based on the results of Drory et al. (2004).  Right
panel: the stellar mass distribution as a function of redshift.  As in
the left panel, the red triangles are galaxies with M$_{*} >$ \hmass, and the 
blue boxes are for galaxies with \mass $<$ M$_{*} <$ \hmass.  The green 
vertical lines show the location of the bins we use in our analysis of the 
statistical properties of these massive galaxies, including their number and 
mass densities. The numbers within each bin label their median redshift.   We
also include as small points in both panels the K-band magnitude and
stellar mass distribution for all galaxies with spectroscopic 
redshifts at $z < 1.4$. }
} \label{sample-figure}
\end{figure*}

The random errors result from Poisson statistics on the number of 
galaxies ($N$) in each redshift and mass bin.  We thus include a component
in our error budget to account for these counting uncertainties.
Stellar mass uncertainties 
are considered in two ways. The first method uses Monte-Carlo simulations of
mass errors to calculate how a typical 0.3 dex error can affect the
measurement of  number densities.  This is the typical error
due to mismeasuring the stellar masses for our sample (e.g., Bundy
et al. 2006).    This factor of two comes from several sources, which
we conclude are all contributing random errors to our mass budget. Below
we examine the reasoning for this, and why we do not include any
systematic stellar mass uncertainties.

It is first important to note that our photometry originates from
SExtractor, and does not use the {\em imcat} software utilised by DEEP2 to
obtain photometry for target selection and luminosity functions (e.g,
Faber et al. 2005).
We investigated systematic issues in the photometry through our
simulations of placing fake galaxies of known magnitude into our images,
and then determining how the detected magnitude differs from the
input.  We found
that when the completeness was high, the retrieved magnitude was
almost nearly identical to the input.  The completeness at 
$K \sim 20$, the faintest magnitude where our galaxies are found,
is $\sim 100$\% (\S 2.1), and we are thus 
unlikely affected by a systematic photometry error. We do not 
include any additional systematic uncertainty into our error budget due
to our detection methods.

We also compare our total SExtractor MAGAUTO magnitudes to magnitudes
measured using 4\arcsec\, dimeter aperture photometry, and find a very slight
change with redshift,
such that we are not detecting all the light in the highest
redshift galaxies using total magnitudes. This however is
a very slight effect, with at most a difference of 0.03 mag which
corresponds to 0.01 dex in stellar mass. Again, however, this effect
is less of an issue for the brightest galaxies, which are the dominate
population in our sample. 

We conclude that the factor of two, or 0.3 dex random uncertainty
in the stellar mass is a representative estimate of the uncertainties
in stellar masses. To determine how a 0.3~dex
random error in stellar masses can change our densities, we 
simulate how the mass function and number densities of our sample
would change after applying this error. We do this by changing 
the measured stellar masses by a random amount, to within 0.3~dex, and then
recalculate the number and mass densities.  The difference
in stellar mass and number densities between the recomputed and original
values is then added to the error budget.

Another major source of error in our number densities originates from cosmic 
variance. Based on the likely bias (b) and dark matter variance 
($\sigma_{\rm DM}$), it 
is possible to calculate the likely cosmic variance (Somerville et al. 2004),
and to include this in our error budget. 
For massive galaxies with M$_{*} >$ \mass, and using the volume of our 
survey within each 
redshift bin, we calculate that the galaxy variance is $\sim 0.3-0.5$. 
This leads to a fractional variation of 20-30\% for the
most massive galaxies. This variance drops by a factor of three for the lower 
mass galaxies in our sample, rendering cosmic variance less of an issue for 
these systems. 
When we consider our four fields, which are widely separated, it reduces the 
estimates of our cosmic variance by a factor of two.  It turns out that
our computed number and mass densities (Table~2) are often dominated
by these cosmic variance errors.

Another source of uncertainty is the fact that if a galaxy does not
have a measured photometric redshift, it cannot be included
in our sample.   As discussed
in \S 3.2, we do not measure photometric redshifts for
galaxies not detected in one of our optical bands. 
To investigate
whether we could be missing significant numbers of massive
galaxies, particularly at high redshift, where we find evolution, we 
examine the number of galaxies
which are not detected in one band, but which could be high redshift
massive galaxies. First, as Figure~3 shows, the K-limit for our
massive galaxy sample is roughly $K = 19$ for M$_{*} >$ \mass galaxies
at redshifts $z < 1.4$.  There are only 16 objects
in our K-selected sample that are within this K-band magnitude limit, 
but are not detected in the B-band, and thus do not have a measured 
photometric redshift.  Most of these galaxies are near the
K-limit with $K > 18$, and none are as bright as the M$_{*} >$ \hmass
galaxies.

These galaxies with missing optical fluxes are unlikely influencing 
our measured number
and mass densities at $z < 1.2$ for the following reasons.  The total
number of galaxies with M$_{*} >$ \hmass at $z \sim 1.4$
is 45. If all 16 galaxies at our K-limit without a photometric redshift
were placed into this bin, it would only increase densities
by 30\%, not enough to account for the non-evolution we see.
Furthermore, a M$_{*} >$ \hmass galaxy with a maximum
M/L ratio would still be brighter than $K = 18$ at 
$z \sim 1.4$, and thus these 16 galaxies are not likely
to be at $z < 1.4$ with a mass M$_{*} >$ \hmass. 
The $10^{11}$~\solm~$<$~M$_{*}~<~10^{11.5}$ \solm bin
at $z \sim 1.4$ contains 725 galaxies, and these
16 galaxies would contribute a neglectable amount to
their number and mass densities at $z \sim 1.4$.  We
therefore conclude that our criteria that galaxies
be detected in all optical bands to be included in
our sample is a very insignificant source of uncertainty
in our analysis at $z < 1.4$.

We are however less complete for
selecting galaxies within our mass ranges at $z > 1.2$. We 
correct for this by adding the number of galaxies at these higher
redshifts that are likely missed. We following Willmer
et al. (2005) for this process, where we consider the likely sampling
based on the K and I magnitudes, and the (I-K) colour for systems
which did not have a measured photometric redshift, and are missing
a flux in one of the two other optical bands (generally the B-band).  We add
into our number and mass densities those objects which have
a optical-NIR colour and magnitude in I and K similar to the distribution
of real galaxies in the given stellar mass and redshift bin. This will 
in fact give an upper limit on the density
corrections, however in reality it makes little difference to the
values.  

\subsubsection{Number and Mass Density Evolution}

One of the most basic methods for understanding the evolution of
massive galaxies is determining how their number densities
and integrated mass densities change
as a function of redshift.  In Bundy et al. (2006) we examined the
overall change in the number densities of massive galaxies, finding
little apparent evolution when comparing mass functions.  
We reexamine this in more detail using our spectroscopic
plus photometric redshift sample. We furthermore examine 
this evolution in smaller redshift bins, and explicitly
state how the evolution of the mass function is occurring,
or not occurring,
for the most massive galaxies, and with what certainly
we can make such statements. 

As we are counting galaxies, as opposed to using the V$_{\rm max}$ method
to calculating mass functions, we differ in our approach
from Bundy et al. (2006). It is
important to therefore test that our methods give consistent results.
We find this is the case using
the published Schechter functions in Bundy
et al. (2006), derived from V$_{\rm max}$ computed mass functions,
and our method of empirically determining
whether there is evolution by counting actual galaxies. We
find an agreement between
the two methods using the Bundy et al. (2006)
redshift bins (0.4-0.7, 0.7-1.0, and 1.0-1.4).  
In the M$_{*} >$ \hmass mass range we find slightly more galaxies
using our counting method.
This is due to the Schechter function under-fitting
the high mass end in Bundy et al. (2006), something
that also occurs for the $z = 0$ mass function fits
(see below).

Figure~4 shows the number and mass density evolution of the combined primary
spectroscopic, and secondary photometric-redshift selected, galaxies 
for systems with masses 
M$_{*} > 10^{11.5}$ \solm and $10^{11}$~\solm~$<$~M$_{*}~<~10^{11.5}$ \solm 
out to $z \sim 2$.    Also included is the 
evolution in number densities for galaxies with masses 
$10^{10.5}$ \solm $<$ M$_{*} < 10^{11}$ \solm.
We show the number densities of galaxies within these mass ranges measured
in the nearby universe out to $z \sim 0.2$ by the 2MASS/2dF galaxy surveys 
(Cole et al. 2001) normalised using the same Chabrier IMF as we use for our
higher redshift comparisons.

It is clear by looking at Figure~4, and analysing the number
densities listed in Table~2, that there is very little 
evolution, statistically, at $z < 1$ for the M$_{*} >$ \lmass systems.
We can conclude that all 
massive M$_{*} >$ \hmass galaxies are present by $z \sim 1$ to within a 
factor of 2-3  (see e.g., Cimatti et al. 2006; Bundy
et al. 2006; Brown et al. 2007).  However, as can be seen in Figure~4 there is
some evolution for these massive galaxies at $z > 1$, with evolution
occurring up to $z \sim 1.5$ as well as up to $z \sim 2$. This is the first
time that a study has had a large enough area to make this claim,
and as we argue below, at least a fraction of the increase in 
total stellar mass seen between $z \sim 1$ and $z \sim 2$ (e.g., Dickinson
et al. 2003) is produced in massive galaxies.

As can be seen by eye, within our observational errors, there is some
evolution in number densities for M$_{*} >$ \mass and perhaps M$_{*} >$ \hmass 
selected galaxies between 
$z \sim 1-1.5$, although it is not clear with a casual examination
whether this is significant. The evolution in the number
and mass densities can be examined quantitatively in a number of ways. 
First, when we consider evolution just within our sample from 
$z = 1.5$ to $z = 1$ we find significant increases at masses M$_{*} <$
\hmass, both in terms of number and mass densities. We describe below
a quantitative analysis of these changes.

Galaxies with stellar masses M$_{*} >$ \hmass show an increase in number 
densities between $z = 1.5$ to 0.4 of a factor of 2.7$^{+1.8}_{-1.7}$. This is
however significant only at the  $< 2 \sigma$ level, considering all 
uncertainties. In fact, all of this apparent evolution occurs at $z > 1$.  
Furthermore, we find a factor of 1.3$^{+0.74}_{-0.53}$ increase in the mass 
density associated with  M$_{*} >$ \hmass galaxies at the same redshift range, 
although this is also at less than 3 $\sigma$
significance. When we investigate evolution from $z \sim 2$ to $z \sim 1$
we find an increase of 11.2$^{+8.7}_{-4.9}$ in number densities, and
a factor of 5.5$^{+4.3}_{-1.7}$ increase in mass densities for systems
with M$_{*} >$ \hmass. This is also an insignificant increase, and
it appears that for the most part galaxies with M$_{*} >$ \hmass
are nearly all formed by $z \sim 2$.  We cannot however rule out a 
factor of 2-3 evolution in number densities for these
galaxies, given our large uncertainties.

We calculate the evolution for M$_{*} >$ \hmass galaxies down
to $z \sim 0$ by comparing with nearby galaxy studies, such as
Cole et al. (2001). It is important to note however that these local
studies were done using different techniques than ours, and there is
potentially significant systematic differences used
to measure the total amount of light in these nearby galaxies, and
in the way the stellar mass was computed.    The largest difference
is that the 2MASS data used to calculate the stellar masses in Cole
et al. (2001) may under-represent the total amount of light, given
its shallow depth, by up to 0.5 mags. As we are potentially also missing
light in our massive galaxies (\S 2.1), a direct comparison between
our data and Cole et al. (2001) is fair, although a systematic difference
of a few 10\%-s of percentage cannot be ruled out.

Through a direct comparison with Cole et al. (2001) there
is a factor of 2.2$^{+1.5}_{-1.4}$ increase in number density,
and a factor of 0.74$^{+0.46}_{-0.31}$ increase in integrated stellar mass
between $z \sim 1.5$ and $z \sim 0$ for M$_{*} >$ \hmass systems. 
These $z \sim 0$ number and mass densities are
taken from the Schechter fits and data presented in Cole et al.
(2001).   Note that the Cole et al. (2001) Schechter function  
under-fits the most massive systems.   We correct this by
explicitly using the number densities for the most massive galaxies
tabulated in Cole et al. (2001).

Galaxies with stellar masses $10^{11}$~\solm~$<$~M$_{*}~<~10^{11.5}$ \solm show
significant evolution in number density.  The number densities
of systems with $10^{11}$ \solm $<$ M$_{*} < 10^{11.5}$ \solm increases
by a factor of 2.2$^{+0.57}_{-0.41}$ between $z = 1.4$ and $z = 0.4$, 
a result significant at $>$ 4 $\sigma$. Just as for the most massive
systems, this evolution occurs completely at $z > 1$.   Similarly, we
find a factor of 2.1$^{+0.6}_{-0.35}$ increase in the integrated mass density
for systems with $10^{11}$~\solm~$<$~M$_{*}~<~10^{11.5}$ \solm
within the same redshift range, also at $ > 4$ $\sigma$ confidence. We find
furthermore, after correcting for incompleteness due to the
R-band limit (\S 4.2.1), that there is a factor 
of 14.5$^{+4.1}_{-2.8}$ evolution in the
number densities, and a factor of 10.7$^{+3.1}_{2.0}$ in mass densities
between $z \sim 2$ and $z \sim 1$ for galaxies with  
$10^{11}$~\solm~$<$~M$_{*}~<~10^{11.5}$ \solm. Both of these are
significant at the 5 $\sigma$ level. 
When we consider evolution from $z \sim 1.5$ to $\sim 0$, using
the comparison to Cole et al. (2001), we find a factor of 
3.0$^{+0.78}_{-0.56}$ increase in number densities, and an increase
of 1.7$^{+0.49}_{-0.29}$ in mass densities.  Both of these increases
are significant at the $> 5$ $\sigma$ level.

These results show that there is some evolution in the
number and mass densities of massive galaxies at $z \sim 1 - 2$.  This
evolution is such that the most massive systems with \mass $<$ M$_{*} <$ \hmass
increase in number and mass densities by factors $> 2-3$ at a significance
$> 3$ $\sigma$.  Taken as a whole, we calculate that the scenario whereby
the stellar mass and number densities of galaxies does not evolve between
$z \sim 1.5$ to $z \sim 0.4$ can be rejected at $> 8$ $\sigma$
confidence. Therefore it does not appear that high mass galaxy formation,
with the exception of M$_{*} >$ \hmass systems,
is complete by $z \sim 1.4$, yet it is largely completed by $z \sim 1$.   
Therefore,
the redshift range $z \sim 1 - 1.5$ is the final epoch for the build up
of the majority of the stellar mass in the most massive galaxies.

\begin{figure*}
 \vbox to 120mm{
\includegraphics[angle=0, width=184mm]{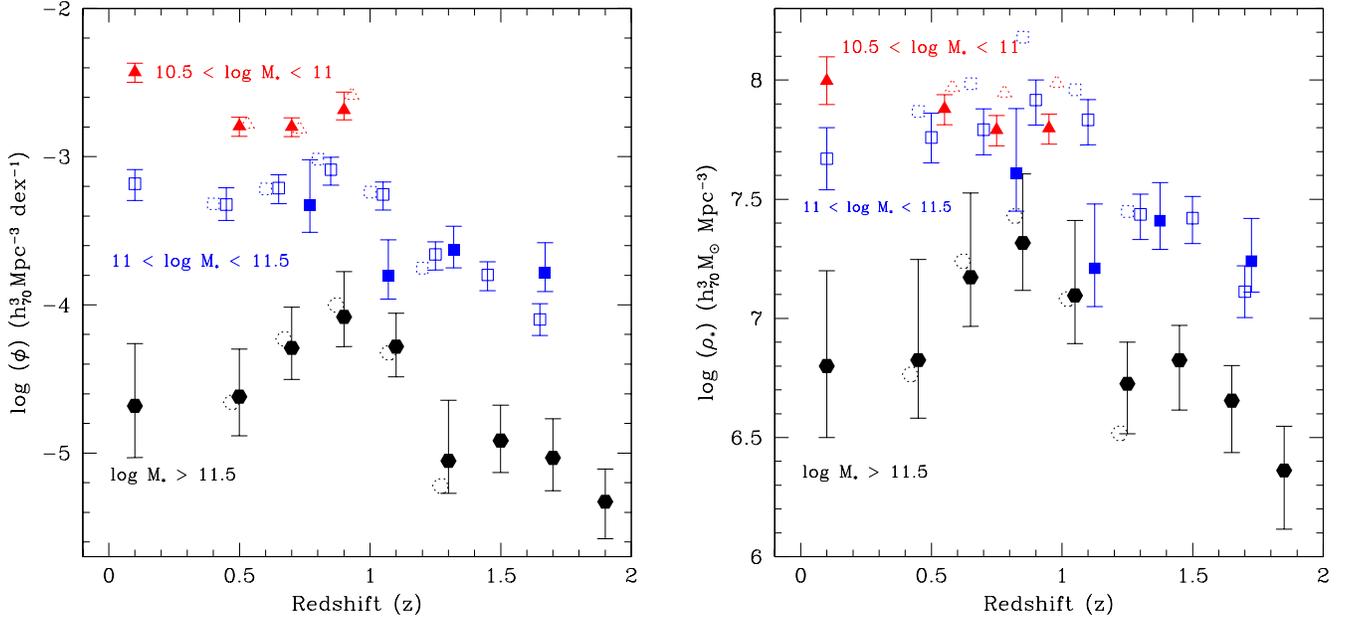}
 \caption{Left panel: the evolution in the number densities for galaxies of
various masses between $z \sim 0.4 - 2$ with the spectroscopic and
photometric redshift samples combined.  Right panel: the
stellar mass density evolution as a function of galaxy mass at the same
redshift intervals.  The points at $z \sim 0$ are taken from Cole
et al. (2001), where the $z \sim 0$ point for the M$_{*} >$ \hmass
galaxies are corrected for the poor fit to these galaxies given in Cole
et al. (2001) (see text).  The error bars listed on both the
numbers and mass densities reflect uncertainties from stellar mass
errors, as well as cosmic variance, and counting statistics.  The
dashed symbols near each data point show how these values would
change if just using photometric redshifts.  The
over-density at $z \sim 0.9$ remains even when we consider just
the photometric redshifts. For comparison we show the mass densities
for systems with \mass$<$ M$_{*} <$ \hmass from Glazebrook et al.
(2004) plotted as solid blue boxes. Note that shifts of $\pm$0.05 in redshifts
have been applied so that data points and errors do not overlap.}
} \label{sample-figure}
\end{figure*}
\vspace{0.4cm}

\subsubsection{Mass Function Peak}

As can be seen in Figure~4, galaxies with 
M$_{*} >$~\hmass have perhaps a surprisingly 
higher galaxy number and mass density at $z \sim 1$ than at $z \sim 0.4$.
This is unlikely a volume effect, as  the total co-moving
volume of our sample at $z \sim 0.9$ is 2.4$\times 10^{6}$ Mpc$^{3}$, while
at $z \sim 0$ the 2dF/2MASS sample occupies 5 $\times 10^{6}$
Mpc$^{3}$ (Cole et al. 2001).  The other redshift bins have
similar co-moving volumes as the $z \sim 0.9$ bin.  Furthermore,
we can see this over-density, to a much lesser degree, in the
number and mass densities using other mass cuts. 

Due to the photometric
selection there are more galaxies at $z > 0.7$ with spectroscopic
redshifts than at lower
redshifts. This should not however create such a difference,
as we are filling in the missing galaxies using photometric 
redshifts.  To test whether this effect is due to the spectroscopic 
redshifts picking up at $z > 0.7$, we
redo Figure~4 using just photometric redshifts. When we do so,
we obtain nearly the same result (Figure~4).  We also see this peak
to a limited degree when only 
examining objects with spectroscopic redshifts.

Another issue is whether the photometric redshifts, most of which
were trained with a neural network method (ANNz), are somehow biased
towards a $z \sim 0.9$ redshift.  We tested this in a number of ways.
First, the spike at $z \sim 0.9$ for the M$_{*} >$ \hmass galaxies
appears to be dominated by fairly low 
number statistics, and the peak is most obvious in Field 2.  
There is perhaps a slight excess in the EGS, but there is no
excess or peak seen in Fields 3 or 4 when examining either the
photometric or spectroscopic redshifts.  Secondly, the ANNz method
may bias the fitted redshifts to exist at only certain values. This is
likely not occurring in our sample for the following reasons.  We are
training our method using only the EGS field, which has a spectroscopic
over-density at $z = 0.7$, while our over-density is found 
at $z \sim 0.9$.  When we examine the distribution of the number of
M$_{*} >$ \mass galaxies as a function of redshift in narrow $\delta$z
= 0.005 bins, we do not find any trend or correlation between 
spectroscopic redshifts and photometric redshifts. That is, the locations
in redshift where photometric redshifts are found are not the same
as the spectroscopic redshifts.

An increase in number density over  the $z \sim 0$ values for massive
galaxies can  be seen, often to an even greater level, in the
mass functions plotted in Bundy et al. (2006), as well
as in the independent MUNICS (Drory et al. 2004) and
COMBO-17 (Borch et al. 2006) fields.
There are a few possibilities, beyond cosmic variance, which can explain
this result.   Although
we conclude that the following are not the causes of the excess, we 
investigate them to show they are unlikely playing a role. The first is that 
the co-moving volumes we utilise to
compute densities depends strongly on cosmological parameters.  The first is 
that if we use a cosmology without a cosmological constant, an open universe
with $\Omega_{\rm m} = 0.3$, the density contrast becomes larger. In fact,
only values higher than $\Omega_{\lambda}$ = 0.7 helps alleviate the problem,
but no reasonable values totally erase the excess.  

Another issue we
examined is whether the measurements of stellar masses are incorrect
due to using the Bruzual \& Charlot models to compute our mass to
light ratios.  Models, such as Maraston (2005) who include a newer
treatement of
thermally pulsing asymptotic giant branch (TP-AGB) stars in their models,
can lower computed mass to light ratios for certain populations (see
\S 3.3). As the TP-AGB 
stars have their most effect at an age of 1-2 Gyr, we investigate the 
fitted ages of our stellar populations to see if they are near these
values.  The average fitted
ages for galaxies at $z > 0.6$ is roughly 4 Gyr for systems
with M$_{*} >$ \mass, and thus it is not likely that the exclusion 
of TP-AGB stars is influencing our mass measurements.  However, fitted
ages using TP-AGB stars might be lower, and in this case, it still
remains possible that our masses are slightly overestimated. However, this
likely cannot account for the entire increase, as the improved inclusion of
the TP-AGB stars would lover the stellar masses by only 20\% (\S 3.3). 
The real cause
of the increase at $z \sim 0.9$ 
is likely cosmic variance, and  even surveys larger than ours will be needed to
probe the very massive end of the mass function with higher certainty.

\subsection{Structures and Morphologies}

Investigating the structures and morphologies of galaxies is becoming 
recognised as one of the most important methods for understanding 
galaxies (e.g., Conselice 2003; Trujillo et al. 2004; Cassata et al. 2005;
Lotz et al. 2006).
Since the morphologies and structures of galaxies have a direct relationship 
to their formation modes (i.e., disks, merging systems, etc), we 
study the 
morphological properties of our sample in some detail.  The overlap of our 
NIR imaging and the ACS imaging in the EGS contains 506 galaxies with 
stellar masses M$_{*}$ $>$ \mass at $z < 1.4$.    For nearly all of
these systems, their magnitudes are bright enough such that effects
due to redshifts will not affect our ability to classify these
systems either by eye, or through quantitative methods (e.g., Conselice
et al. 2000c; Windhorst et al. 2002; Papovich et al. 2003, 2005; 
Taylor-Mager et al. 2006).  Note that the following morphological
and structural analyses are only within the ACS coverage of the
EGS field.

\subsubsection{Visual/Classical Morphologies}

We study the structures and morphologies of our sample within the
EGS using two different
methods.   The first is a simple visual estimate of morphologies based
on the appearance of our galaxies in the ACS imaging.  While the ACS
imaging of the EGS field includes both F606W and F814W imaging, we only
use the F814W band imaging in our structural/morphological analyses.
The outline of our classification
process is given in Conselice et al. (2005a).  Basically, we 
place each galaxy into one of nine categories: compact, elliptical, lenticular (S0),
early-type disk,  late-type disk, edge-on disk, irregular, merger/peculiar, 
and unknown/too-faint. While classifying these galaxies by eye it became
apparent that many of the early-types appear to have a slight distortion.
We therefore added a sub-class of distorted ellipticals to our classifications.
These classifications are very simple, and are only 
based on appearance. No other information, such as colour or redshift, was
used to determine these types. A short explanation of these types is 
provided below, with the number we find in each class listed at the end 
of each description.

\begin{enumerate}

 \item Ellipticals (E): Ellipticals are centrally concentrated galaxies with 
no evidence 
for lower surface brightness, outer structures.  (263 systems, 68 of
which are classified as disturbed Es)

 \item Lenticular (S0): A galaxy is classified as an S0 if it
appears elliptical-like, but contained a disk-like outer structure with 
no evidence for spiral arms, or clumpy star forming knots, or other
asymmetries. (23 systems)

 \item Compact - A galaxy is classified as compact if its structure
is resolved, but still appears compact without any substructure. It is 
similar to the elliptical classification in that
a system must be very smooth and symmetric. A compact galaxy differs
from an elliptical in that it contains no obvious features such as an extended
light distribution or envelope.  (65 systems)
     
 \item Early-type disks: If a galaxy contains a central 
concentration with some evidence for lower surface brightness outer 
light in the form of spiral arms or a disk, it is classified as an 
early-type disk. (28 systems)
 
 \item Late-type disks: Late-type disks are galaxies that appear to 
have more outer low surface brightness disk light than inner concentrated 
light. (18 systems)

 \item Edge-on disks: disk systems seen edge-on and whose face-on morphology 
cannot be determined, but is presumably an S0 or spiral. (12 systems)
   
 \item Irregulars: Irregulars are galaxies that appear to have no 
central light concentration, and a diffuse structure, sometimes with 
clumpy material present. (2 systems)
    
 \item Peculiars: Peculiars are systems that appear to be disturbed,
or peculiar looking, including elongated/tailed sources. These galaxies are 
possibly in some phase of a merger (Conselice et al. 2003a), and are more
common at high redshifts. (92 systems)
   
 \item Unknown/too-faint: If a galaxy is too faint for any reliable 
classification it was placed in this category. Often these galaxies appear 
as smudges without any structure. These could be disks or ellipticals, but 
their extreme faintness precludes a reliable classification. (3 systems)
\end{enumerate}

We repeat our visual classifications for a fraction of our sample, finding
an initial misclassification at the 5\% level.  We also compared our
eye-ball estimates to the CAS quantitative values discussed in
\S 4.3.4, which allowed us to identify a significant fraction of the
misclassified systems (see Appendix A), which were then corrected
to their proper morphological type.  We only perform these classifications
out to $z \sim 1.4$ (cf. higher redshifts in Conselice et al. 2007a), and
only use the F814W band for the classifications.  This presents a
potential k-correction issue, as the F814W band reveals bluer light
at higher redshifts.  However, through tests with the F606W band at
redshifts where F814W probes well into the rest-frame optial,
as well as tests on classifications in the HDF and Hubble Ultra 
Deep Field, show that morphological typing within the F814W band gives
very similar results, with at most a 5\% difference between
rest-frame B-band and the observed I-band at $z < 1.4$ 
(Conselice et al. 2005a).  

When we discuss classifications in terms of early/mid/late types, these
refer to: early (E/Compact/S0), mid (Sa-Sb), late (Sc-Irr).
For the most part we find from our visual analysis that most of
the $z < 1.4$, and M$_{*} >$ \mass systems, in which we can see structure,
have an early-type morphology.    There is however a small,
but significant, morphological diversity among the M$_{*} >$ \mass 
galaxies.  We find that over all redshifts, 69\% of the M$_{*}$ $>$ \mass 
systems are 
early-types (elliptical, S0, compact), while 10\% are disks, and 18\% are 
peculiars. This is perhaps a surprisingly high fraction of peculiars within our
sample, and suggests that some of these systems are still undergoing
some type of mass assembly, possibly through merging or star formation.  
This however changes
when we examine only the most massive systems with M$_{*} >$ \hmass.
These galaxies are $\sim 80-90$\% early-type over all redshifts, with a roughly 
similar number of mergers and disks making up the remainder.  These results
remain essentially the same to within 5\%, if we consider the Eddington bias 
bringing lower mass galaxies into our mass cuts due to observational 
uncertainty.
This confirms and expands with a larger sample, earlier work on massive
galaxies which also concluded that more massive galaxies appear to
develop an early-type morphology before lower mass systems (Brinchmann
\& Ellis 2000; Bundy et al. 2005).

\begin{figure*}
 \vbox to 120mm{
\includegraphics[angle=0, width=154mm]{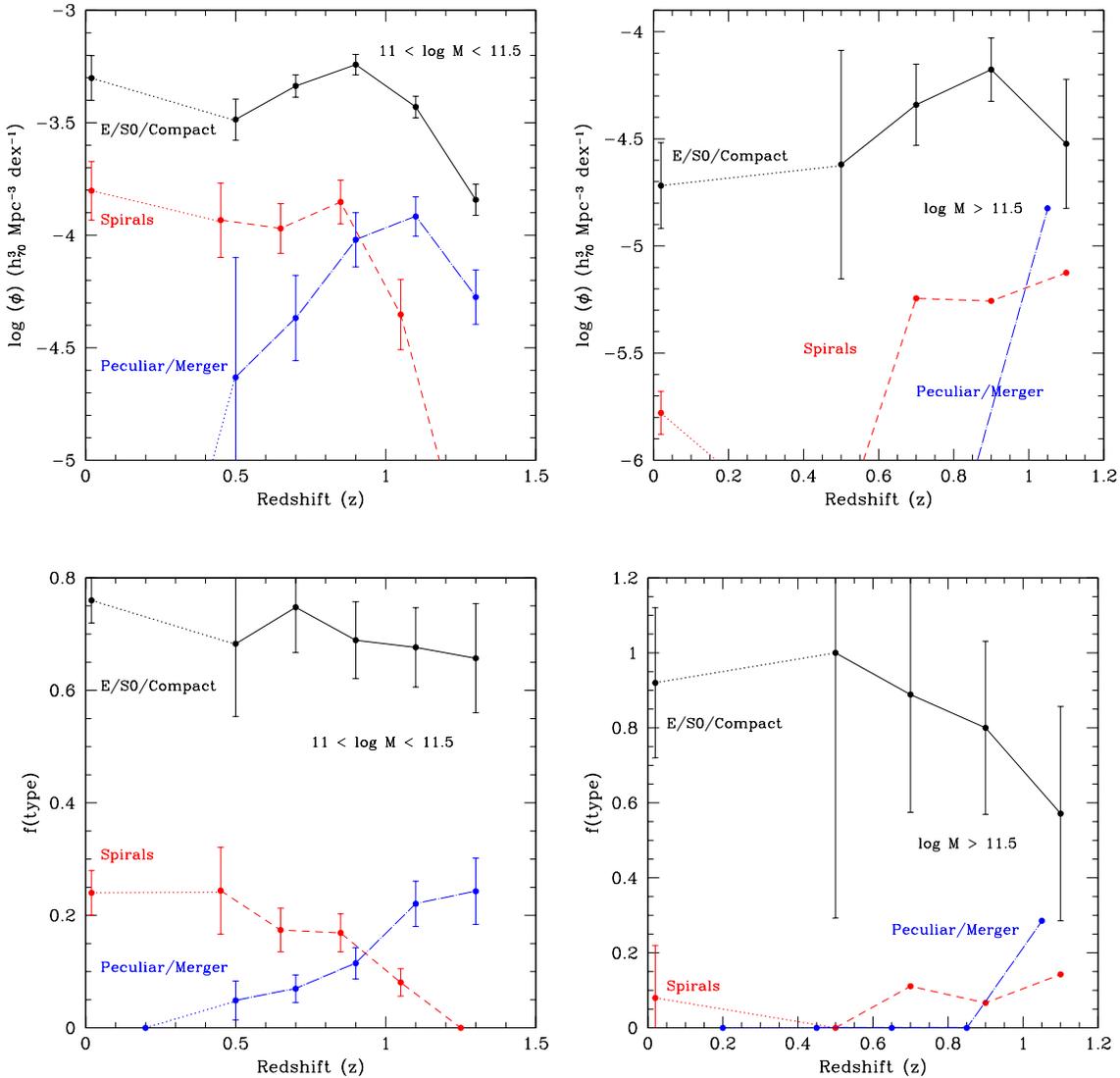}
 \caption{Visual estimates of the morphological evolution for
the most massive galaxies within the EGS portion of our survey, where
we have ACS data from Hubble.  This figure is divided into two mass
ranges and shows in the upper panels the evolution of galaxy number
densities at $z < 1.4$, while the bottom panel shows the evolution
in the relative numbers of different morphological types selected by
mass.   The left panels shows the morphological break down up to 
$z \sim 1.4$
for systems with \mass $<$ M$_{*} <$ \hmass.  The right hand side shows the 
corresponding
evolution for galaxies with M$_{*} >$ \hmass.  The error bars on the
M$_{*} >$ \hmass galaxies are significantly higher than for the  
\mass $<$ M$_{*} <$ \hmass galaxies due to
the smaller number of galaxies, and
are only plotted for the elliptical/S0s, although the peculiars/mergers
and spirals have similar errors.  Both plots demonstrate that ellipticals
dominate the population up to $z \sim 1.4$. Interestingly, it also appears
that the fraction of spirals increases from high to low redshift,
while peculiars decline from higher to lower redshifts.  The morphological
fractions and densities form the $z = 0$ systems are taken from
Conselice (2006a).}
} \label{sample-figure}
\vspace{9cm}
\end{figure*} 

In Figure~5 we plot the visual estimates for
how the morphological break-down for both the M$_{*}$ $>$ \hmass and 
\mass $<$ M$_{*} <$ \hmass galaxies evolves
from $z \sim 1.4$ to 0.1. The $z \sim 0$ points are based on morphologies
from the RC3 catalogue, as described in Conselice (2006a).   The most
remarkable aspect of this evolution is that the fraction of elliptical
galaxies with masses greater than M$_{*} =$ \mass is relatively constant out
to $z \sim 1.4$, at about 70\%.  There is perhaps a slight increase for
the \mass $<$ M$_{*} <$ \hmass systems at $z = 1 - 1.5$, as seen in
previous work (Bundy et al. 2005).
Remarkably, there seems to be a gradual
trend for the remaining 30\% to transition from mergers/peculiars into disks.
It is interesting that such massive disk galaxies exist at these redshifts,
which must have evolved into the M$_{*} >$ \mass bin through star formation, or
through a morphological transformation from an early-type to a spiral 
(\S 4.3.2).

In summary, what we find is that the majority of massive galaxies up
to $z \sim 1.4$ have an early-type morphology (see also Bundy
et al. 2005). This is consistent
with the findings that most of these systems are already formed
in terms of their stellar masses at similar epochs (\S 4.2).  However,
it is clear that $\sim 30$\% of galaxies with M$_{*} >$ \mass
are not early-types, which suggests that there is evolution in the
massive galaxy population that cannot be seen simply through the evolution
in number and mass densities.  The disk galaxies we see suggest
that there is some star formation occurring, and the peculiars
suggest that there is  merger activity. We investigate both of
these galaxy-types, and what they imply for the evolution of
the most massive galaxies, in the next sections.

\subsubsection{Massive Disk Galaxies}

As can be seen from Figure~5, there is a significant, $> 4 \sigma$, increase 
in the number, and relative fraction, of massive M$_* >$ \mass disk galaxies 
at $z < 1.4$.  This result is robust to morphological k-corrections (\S 4.3.1;
Conselice et al. 2005), and shows a real increase in massive disk galaxies
through time. We investigate the possible reasons for this
increase in this section.

As already noted, there is a gradual transition for the non-ellipticals
with M$_{*} >$ \mass to appear as peculiars at $z > 1$, and as disks
at $z < 1$.  What is likely occurring is that the
higher redshift peculiars are transforming into ellipticals after they 
dynamically relax.  This leaves the problem of why are there are significantly 
more disks at lower redshifts, and why the fraction of early-types does not 
change with redshift. That is, if the highest redshift peculiars are 
becoming ellipticals, we would expect the early-type fraction to increase at 
lower redshift, but it remains roughly constant.
One possible scenario is that some of the already massive ellipticals 
acquire a disk through accretion of intergalactic gas, and the accretion of 
low-mass satellite galaxies (e.g., Abadi et al. 2003).  Although candidates
for this disk formation are seen at slightly higher redshifts both
morphologically and through kinematics
(e.g., Erb et al. 2003; Conselice et al. 2004; Labbe et al. 2004; 
Elmegreen et al. 2005; Forster-Schreiber et al. 2006; Genzel et al. 2006; 
Bouche et al. 2007), it is unlikely that these large disks without massive 
bulges are formed in this manner.

The rise of these massive disks are explainable
by systems with disks masses M$_{*} <$ \mass at $z > 1$ 
being brought up into the M$_{*} >$ \mass bin through
star formation or minor mergers.   As we have seen earlier, the number
densities of galaxies with M$_{*} >$ \mass increases slightly
with time. Therefore these disk galaxies were likely lower mass 
(M$_{*} <$ \mass) systems, and achieved M$_{*} >$ \mass status by 
$z \sim 0.5$ through star formation or minor merger accretion 
activity (\S 5.1). For structural reasons this seems like the most likely
scenario, and can explain the nearly constant elliptical fraction,
despite the morphological transformation of peculiars, by
simply increasing the total number of $>$ \mass galaxies by adding new
spirals.  

This idea is backed up by the evolution in the number densities, and
relative number fractions, of the different morphological types shown
in Figure~5.  The total number densities of disk galaxies at $z < 1$
is higher than the number densities of peculiar galaxies at $z > 1$. 
This means that it is difficult for the peculiars, without significant 
new star formation, to produce all of the spirals seen at lower redshifts. At
the same time the peculiars are decreasing in number density
at $1 < z < 1.4$, the densities of the ellipticals are increasing. This
causal relationship is highly suggestive that these peculiars are
transforming morphologically into early types by $z \sim 1$.  After
$z \sim 1$ there are very few peculiars in the mass range
\mass $<$ M$_{*} <$ \hmass, and no further growth in the number densities
of early types.

Another argument against a direct peculiar to spiral transition is
the structures of these massive disks. The nine most massive spirals/disks
 in our sample are shown in Figure~6.  As can be seen, these systems are 
often dominated by disk light, and do not appear to have particularly 
large bulges.  Some systems also appear to have rings and bars.
Because these spirals have such large disks and small bulges, it is unlikely 
that the peculiar galaxies at $z \sim 1.3$ are transforming into these 
systems. If this were the case, we would expect the spirals to show more 
disturbances, or at least a large bulge formed through the merger.   
Massive disk galaxies in previous work are also found to be largely in place
in terms of sizes, morphologies, and masses by $z \sim 1$ 
(Ravindranath et al. 2004; Jogee et al. 2004; Conselice  et al. 2005b), thus 
any evolution in disks must be gradual.  An interesting future study would
be to study in detail these high mass disk galaxies at $z < 0.5$ to
decipher their origin in more detail.

\begin{figure}
\hspace{0cm}
 \vbox to 120mm{
\includegraphics[angle=0, width=80mm]{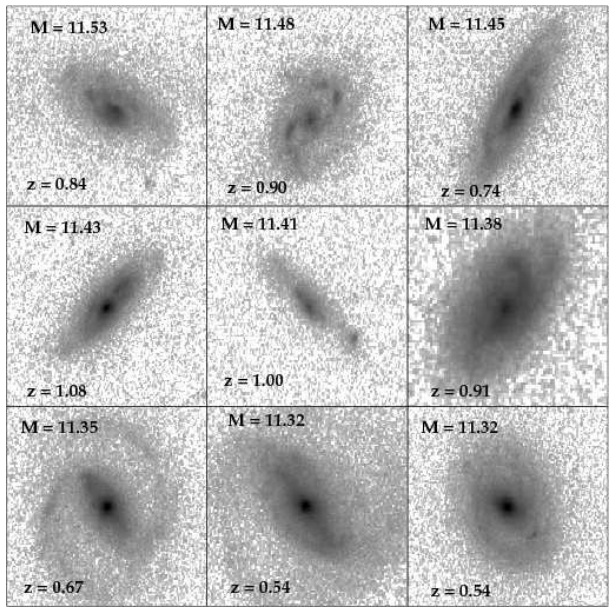}
 \caption{The most massive disk galaxies in our sample with the value
of their stellar mass (in log M units), and
their redshifts labelled. These disks have few tidal distortions
due to interactions or minor mergers, and they all appear to lack
large bulges. They also appear largely
devoid of star formation, and occasionally they have rings and
bars.}
} \label{sample-figure}
\end{figure}

\subsubsection{Disturbed Ellipticals}

It is important to note that although early-types (classified E/S0/compact) 
dominate the galaxy population at both the M$_{*} >$ \mass and  
M$_{*} >$ \hmass mass selection limits, these 
galaxies often contain evidence for morphological peculiarities. 
Usually these are in the form of outer low surface brightness features, or
multiple cores.  These are most likely the result of recent past
merger activity in these galaxies, either through major or minor mergers.  
Examples of these distributed ellipticals are shown in Figure~7. 

A total of 68 out of 263 (26$\pm3$\%) of the ellipticals with M$_{*} >$ 
\mass show some
internal substructure visible by eye (Figure~7). As Figure~8 shows,
the fraction of ellipticals with these distorted structures is relatively
constant over time.  There is perhaps a slight increase in this fraction
at lower
redshifts, which if real can be accounted for by surface brightness dimming.
Peculiar ellipticals have been seen in other ways previously,  such as 
through colour gradients and colour structures in ellipticals (e.g., Menanteau
et al. 2005;  Stanford et al. 2004; Teplitz et al. 2006), resulting from 
star formation and which may be related to the features seen here. These 
previous studies have generally found that the lower mass ellipticals 
contain these star formation signatures.  These
morphological disturbances however do not appear more common in the lower
mass ellipticals, and in fact, 36\% of the M$_{*} >$ \hmass ellipticals
show this signature - a higher fraction that for the \mass $<$ M$_{*} <$ 
\hmass  population.
If these features are formed from a merger of various types, then it
would be an anti-downsizing signature, suggesting that many giant
ellipticals are still forming at $z < 1.4$, perhaps through dry-merging.

\subsubsection{CAS Structural Analysis}

We use the CAS (concentration,
asymmetry, clumpiness) parameters to probe the structures of our galaxies
quantitatively.  The CAS parameters are a non-parametric method for
measuring the structures of galaxies on resolved CCD images (e.g.,
Conselice et al. 2000a; Bershady et al. 2000; Conselice et al. 2002; 
Conselice 2003).  Our main 
purpose in using the CAS system is to identify relaxed massive
ellipticals, as well as any galaxies that are involved in recent
merger activity.  The
basic idea is that galaxies have light distributions
that reveal their past and present formation modes (Conselice 2003). 
Furthermore, well-known galaxy types in the nearby universe fall in 
well defined regions of the CAS parameter space.  We can then automatically 
determine the structures of galaxies and
classify them according to where they fall in this space (Conselice
2003).     For example, the selection $A > 0.35$ locates systems
which are nearly all major galaxy mergers (Conselice et al. 2000b; 
Conselice 2003; Hernandez-Toledo
et al. 2006; Conselice 2006b); although `dry' mergers will not be
detected easily through this method (Hernandez-Toledo et al. 2006).

We apply a revised CAS system to our massive galaxy sample
to determine their structural parameters.  There are two caveats to 
using the HST one orbit ACS imaging of these galaxies.  
The first is that there are 
morphological k-correction and surface brightness dimming effects which will 
change the measured parameters, such that the 
asymmetry and clumpiness indices will decrease (Conselice et al. 2000; 
Conselice 2003), and the concentration index will be less reliable (Conselice
2003).  There is also the issue that systems at $z > 1.2$ are viewed in 
their rest-frame ultraviolet using ACS data, which
means that there are complications when comparing their measured structures
with the calibrated rest-frame optical indices for nearby galaxies.

Figure~9
shows the concentration/asymmetry and asymmetry/clumpiness projections of
the CAS plane.  As can be seen in the CA plane,  the early-type galaxies 
mostly fall into the $z \sim 0$ calibrated early or mid-type region, and 
disks are found in the mid-type and late-type regions.  Interestingly, the 
distorted (as selected by-eye) ellipticals are significantly 
more asymmetric than the normal ellipticals.

\begin{figure}
 \vbox to 120mm{
\includegraphics[angle=0, width=90mm]{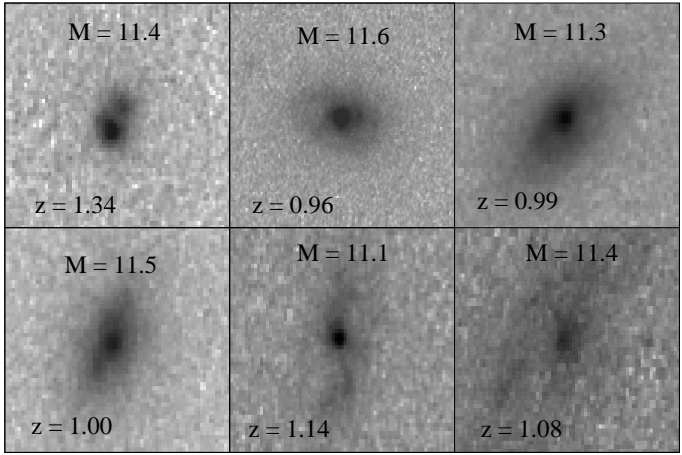}
 \caption{Images of six galaxies classified as distorted
ellipticals. These are among the most peculiar of our sample, with
most peculiarities not easily reproduced on paper, but easily seen
with a viewer, and within the quantitative CAS parameters.  Listed
on each panel is the value of log M for each galaxy,
as well as their redshift.}
} \label{sample-figure}
\end{figure}

The CAS approach for classifying these galaxies is roughly similar to what
the visual morphologies suggest, although there are important
exceptions.  Details of the differences between these two methods are further 
elaborated on in Appendix~A.   For the most part, galaxy types, as determined by eye, are 
where they are expected to be found on this diagram.
One important exception is that many of the visually classified non-distorted 
early-types are not located in the corresponding $z \sim 0$ part of the 
CAS diagrams.    While  these
systems have concentration indices similar to nearby giant 
ellipticals (Conselice 2003), they often
have higher than expected asymmetry indices.  An examination of these
galaxies, and the residuals from a rotation and subtraction reveals that many
of these systems have internal structures, likely resulting from mergers
of various forms, that cannot be easily seen by eye.
 
Furthermore, the SA plane (Figure~9) shows an interesting pattern, such that 
most of our
galaxies appear to deviate from the $z \sim 0$ relation, shown by the solid 
line. This deviation is such that all galaxies are 
more asymmetric than their $z \sim 0$ counterparts, although the slope of
the relationship between asymmetry and clumpiness is similar to that
found at $z \sim 0$.  This implies that galaxies
of a given morphological type, particularly the early-types, have 
a higher degree of bulk asymmetry in their structures.  The concentration
and clumpiness values, which trace to first order the stellar mass and
star formation, remain similar to what we expect from samples
in the nearby universe. Finally, it
appears that the most asymmetric systems, and those that deviate the most
from the A-S relationship are the peculiar/merging systems, or those
ellipticals that show some evidence for structure due to recent tidal
events.  This shows that our visual estimates of merging
systems is borne out by this quantitative analysis.

\subsubsection{Merger Fractions from CAS parameters}

One of the great benefits of using the CAS system for
finding mergers is that it allows us to quantify the
merger fraction and merger rate, and the number of mergers occurring in
a galaxy population (Conselice et al. 2003a; Conselice 2006).

The first observation we can derive from our
CAS values is the evolution of the merger fraction. We determine the
merger fraction for the M$_{*} >$ \mass galaxies using
the criteria, outlined in Conselice (2006) of,

\begin{equation}
{\rm A > 0.35,\, A > S.}
\end{equation}

\noindent Using these criteria, we determine the merger fraction 
for the M$_{*} >$ \mass galaxies in our sample (Figure~8).
As can be seen, there is a slight increase with redshift in the 
merger fraction such that it increases as $(1+z)^{1.3}$, similar
to the evolution seen in lower mass galaxies (Conselice et al. 2003; 
Bridge et al. 2007) up  to $z \sim 1$.

By using the number densities for our systems, and the merger
time-scales for our CAS method, we can calculate
the merger rate for our M$_{*} >$ \mass galaxy sample.
The number densities for our systems are taken
directly from Table~2 in this paper, and the time-scales 
for merging
are derived from equation 10 in Conselice (2006), based
on N-body models analysed using the CAS approach.  From
this, we derive a merger CAS time-scale of $\tau_{\rm} = 
0.43\pm0.05$ Gyrs for a galaxy with a mass of
3$\times 10^{11}$ \solm. Note that this time-scale is not the 
total merger time, but the time-scale
in which the CAS system would identify an ongoing merger. 
Details for how this time-scale is computed are included
in Conselice (2006).

The merger rate for our systems can be calculated through
the merger rate equation,

\begin{figure}
 \vbox to 120mm{
\includegraphics[angle=0, width=90mm]{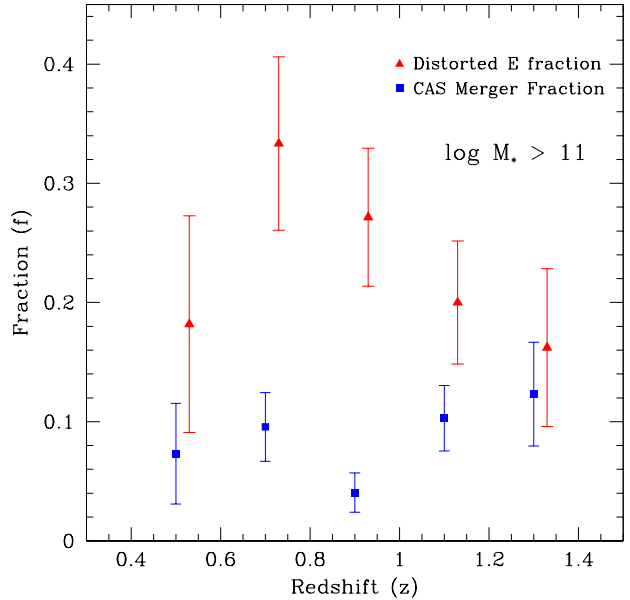}
 \caption{The evolution from $z \sim 1.4$ to 0.4 
of the merger fraction derived
through the CAS approach for galaxies with M$_{*} >$ \mass.
We find that the merger fraction as derived with the CAS
parameters declines as $(1+z)^{1.3}$.  Also plotted is the fraction of early
types which are distorted within the same mass range
of  M$_{*} >$ \mass.  Note that like all of the structural and
morphological plots, these data are only taken from the EGS.}
} \label{sample-figure}
\end{figure}

\begin{equation}
{\rm \Re(z) = f_{m}(z) \cdot \tau_{m}^{-1} n_{m}(z)}
\end{equation}

\noindent where n$_{\rm m}$ is the number densities of
objects, and f$_{\rm m}$ is the merger fraction. Note that
this is not the galaxy merger fraction, which is
the fraction of galaxies merging, which is roughly 
double the merger fraction (Conselice 2006).  We
find that statistically the merger rate for these
M$_{*} >$ \mass galaxies is constant from 
$z \sim 0.4 - 1.4$, and is on average
log $<\Re ({\rm Gyr^{-1} Gpc^{-3}})> = 4.3^{+0.4}_{-0.7}$.

We can furthermore calculate the total number of major
mergers a galaxy with a stellar mass M$_{*} >$ \mass undergoes
from $z \sim 1.4$ to $z \sim 0.4$ using equation (11)
in Conselice (2006).  We calculate that the average
number of mergers a massive galaxy with M$_{*} >$ \mass
will undergo from $z \sim 1.4$ to 0.4 is N$_{\rm m} = 
0.9^{+0.7}_{-0.5}$.  Thus, on average a massive
galaxy will undergo about one major merger from $z \sim 1.4$
to $0.4$, roughly consistent with previous results
(Conselice 2006; Bell et al. 2006).

\subsection{Star Forming Properties of Massive Galaxies}

\begin{figure*}
 \vbox to 150mm{
\includegraphics[angle=0, width=154mm]{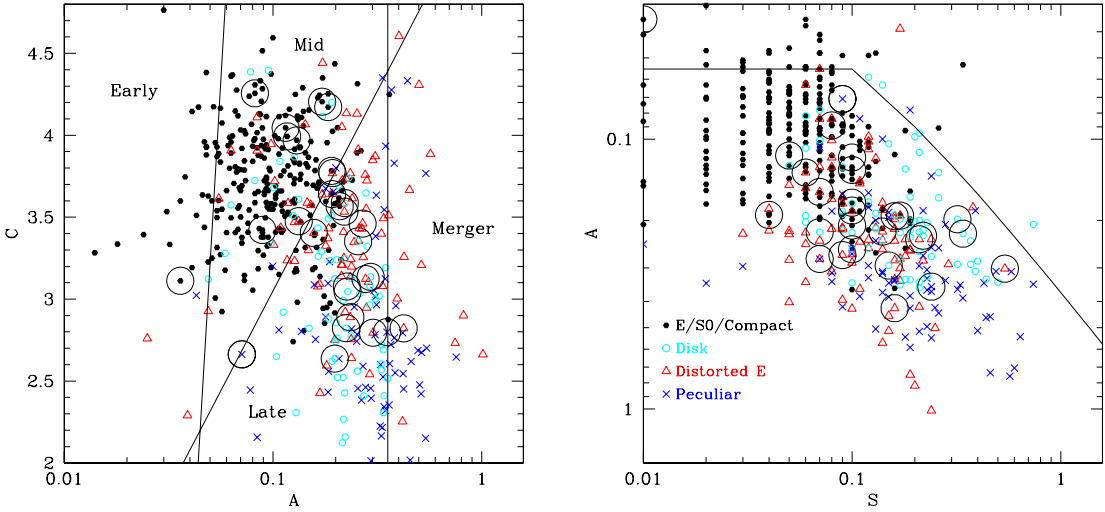}
 \caption{The concentration-asymmetry, and asymmetry-clumpiness projections 
of CAS space as occupied by galaxies with M$_{*}>$ \mass.  Points on 
this diagram are ellipticals/S0/compact (black solid circles), peculiar
ellipticals (red triangles), peculiars/mergers (blue crosses), and disks (cyan open circles). Circled objects are those detected in the X-ray by Chandra.
The lines denoting different galaxy types are from a $z \sim 0$ calibration
described in Conselice (2003). The solid line in the A-S projection shows the
relationship between these two parameters in the nearby universe for normal
non-merging galaxies.  Note that few of the visually classified normal
ellipticals appear in the corresponding part of CAS parameter space due to 
their slightly high 
asymmetries. This can also been seen for all galaxies in the 
clumpiness-asymmetry diagram, showing that most of these asymmetries are 
produced by large-scale features, and not star formation.}
} \vspace{-5cm} \label{sample-figure}
\end{figure*}

\subsubsection{General Trends on the Colour-Magnitude Diagram}

A major question concerning high-mass galaxies is whether
or not these systems have ongoing star formation at high
redshifts. While it is commonly thought that massive
galaxies have finished their assembly and star formation by 
$z \sim 1$, our results suggest otherwise.  This is not the 
first claim for this, as evidence has accumulated during the last few
years that ellipticals, including massive ellipticals, tend to
have some ongoing star formation at $z \sim 1$ (e.g.,
Stanford et al. 2004; Teplitz et al. 2006) and even at
$z \sim 0$ (e.g., Donas et al. 2006).  Star formation, as measured
through blue colours, found in 
morphologically selected early-types, tends to be found in 
lower mass systems (Bundy et al. 2005).  
We therefore investigate the fraction of massive
galaxies undergoing star formation during this time,
and the amount of stellar mass added due to this star formation.

\begin{figure*}
 \vbox to 150mm{
\includegraphics[angle=0, width=154mm]{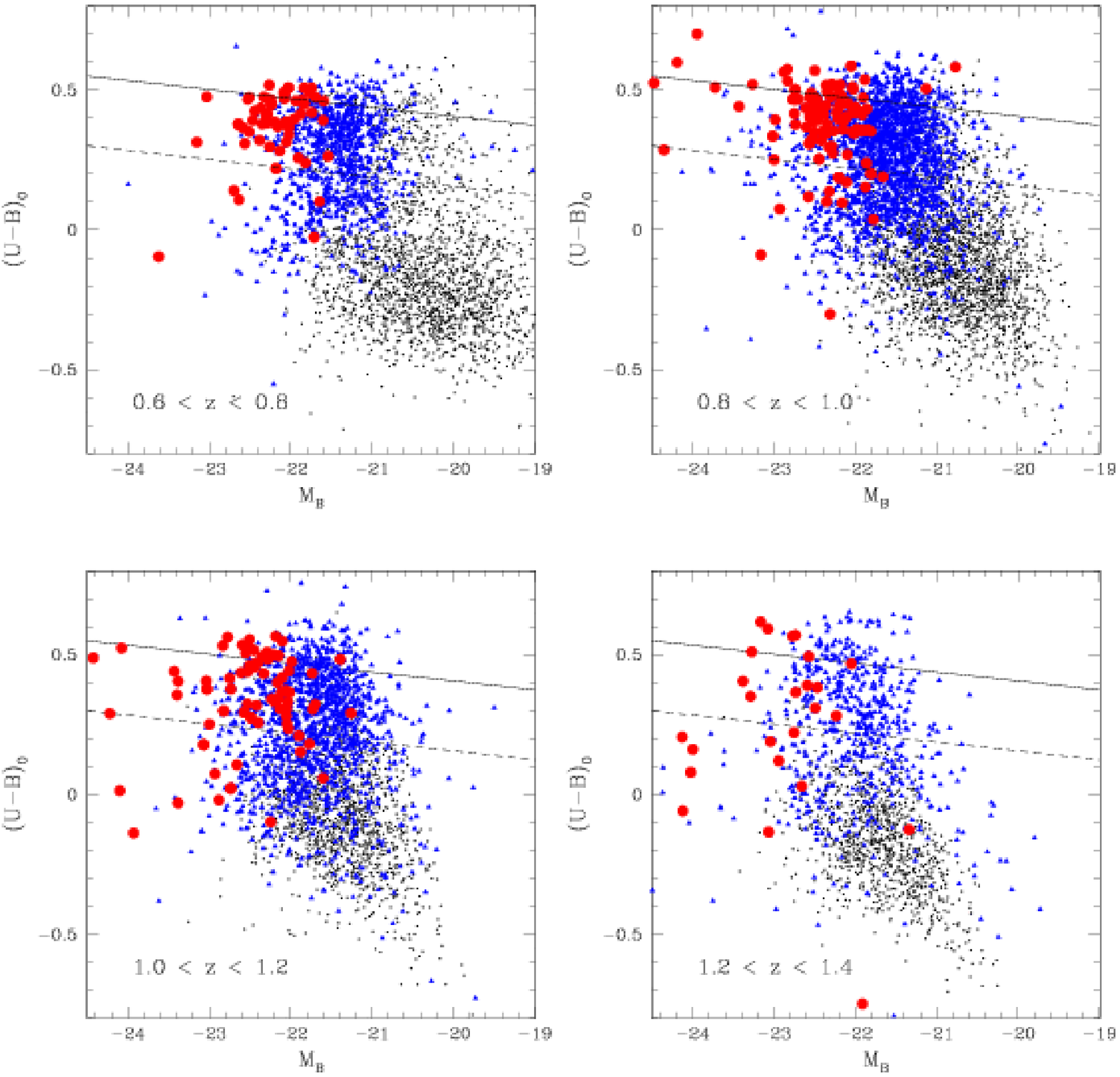}
\caption{The $(U-B)_{0}$ vs. M$_{\rm B}$ diagram for galaxies with
M$_{*} > 10^{10.5}$~\solm from $z = 0.6$ to $z = 1.4$. Included
are both the spectroscopic and photometric redshift samples.  The
large red points on each panel are for those galaxies with 
M$_{*} > 10^{11.5}$~\solm. The blue triangles show
the location of the systems with $10^{11}$ \solm $<$ M$_{*} < 10^{11.5}$ \solm.
The solid line in each diagram
shows the location of the red-sequence as defined in
Faber et al. (2005), and the dashed line is the demarcation
between red and blue galaxies.}

} \vspace{2.5cm} \label{sample-figure}
\end{figure*}

We begin our examination of the star forming properties of 
M$_{*} > 10^{11}$ \solm galaxies by 
examining their location on the rest-frame B-band absolute magnitude
(M$_{\rm B}$) vs. rest-frame colour ($(U-B)_{0}$) 
diagram (Figure~10). These quantities are inferred from the observed
magnitudes through model fitting, utilising the redshift and a series
of k-corrections (Willmer et al. 2006).  
Galaxies appear to separate into a red-sequence and a 
blue cloud 
in this parameter space (e.g., Strateva et al. 2001; Im et al. 2002; 
Baldry et al. 2004; Bell et al. 2004; 
Faber et al. 2005).   The evolution of galaxies on the colour-magnitude
diagram has potential meaning for understanding how
galaxy evolution occurs.  In Figure~10 we plot the $(U-B)_{0}$
vs. M$_{\rm B}$ diagram for galaxies in our fields, with the most
massive galaxies with M$_{*} > 10^{11.5}$ \solm and 
$10^{11}$ \solm $<$ M$_{*} < 10^{11.5}$ \solm labelled.  The 
dividing line between red-sequence and blue cloud galaxies is shown 
as the dotted line in Figure~10, taken from Faber et al. (2005);

\begin{equation}
(U-B)_{0} = -0.032\times(M_{B} + 21.52) + 0.454 - 0.25.
\end{equation}

\noindent We find at the highest redshift bin,
$1.2 < z < 1.4$, a significant number of
massive galaxies that are not on the red-sequence.  In
general, the number of massive galaxies on the red-sequence 
increases at lower redshifts.   This is also true when
we plot these galaxies in terms of their morphologies, where we find
that many early-type galaxies are not on the red-sequence.

The quantification of the fractional 
evolution of galaxies on the red-sequence is displayed in
Figure~11. This shows that the most massive systems 
with M$_{*} > 10^{11.5}$ \solm\,
generally fall in the red-sequence region at
all redshifts, but with a significant number of systems in the blue
cloud region at $z > 0.8$. As Figures~10 and 11 demonstrate,
the fraction of galaxies on the red-sequence, with colours redder
than the value given in eq. (3), increases with time at all
masses.   At the highest redshifts
we can probe star formation, at $z \sim 1.3$, about 60\% of the 
M$_{*} > 10^{11.5}$ \solm\
galaxies are on the red-sequence, yet this fraction increases
to 100$^{+0}_{-26}$\% by $z \sim 0.4$.  Galaxies with lower
masses show a similar pattern, yet lower mass galaxies
always have a lower fraction of galaxies on the red-sequence
at all redshifts, up to $z \sim 1.4$. 

This also leads to a very important conclusion regarding the
red-sequence and high mass galaxies. Previous studies have
examined the increase of the amount of stellar mass on the red-sequence,
finding as much as a factor of two increase 
since $z \sim 1$ (Bell et al. 2004; Faber et al. 2005). However,
this increase is due to galaxies appearing on the red-sequence,
which were previously blue, and not due to in-situ growth on
the red-sequence itself. This can be clearly seen in Figure~11 where
the massive galaxies are gradually moving onto the red-sequence with
time, which can also be seen in the decline in the number of
blue massive galaxies found in the universe since $z \sim 1$. 
This is not consistent with the idea that
the red-sequence grows solely through the so-called `dry mergers'.
Although merging may be present within the red-sequence, and within our
massive galaxy sample, it does not appear to be the dominate
method whereby the red-sequence grows.  

\begin{figure}
\includegraphics[width=84mm]{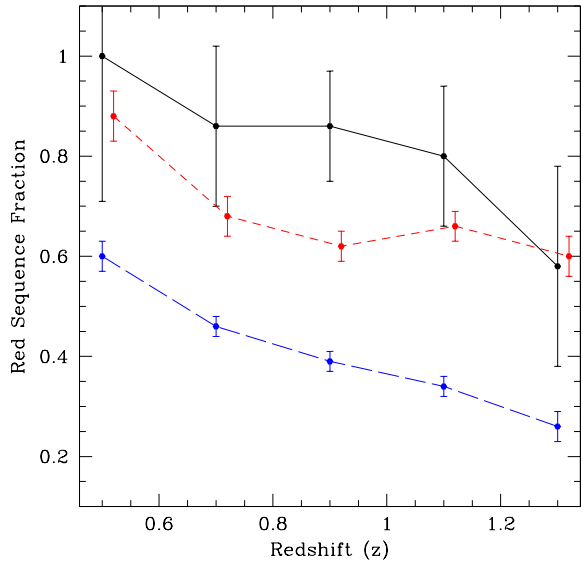}
 \caption{The fraction of galaxies of various masses which are on the
red-sequence as defined in \S 4.4.1.  The solid upper line shows the
evolution of the fraction of galaxies with M$_{*} >$ \hmass which
are on the red-sequence, while the red short-dashed and blue 
long-dashed lines
show the evolution for systems with $10^{11} <$ 
M$_{*} < 10^{11.5}$ \solm, and $10^{10.5}$ \solm $<$ M$_{*} < 10^{11}$ \solm.}
 \label{sample-figure}
\end{figure}

\subsubsection{Quantification of Star Formation}

We quantify the star formation in our sample of massive galaxies using
[OII] line equivalent widths from DEEP2 spectroscopy, 
and through MIPS 24$\mu$m fluxes from Spitzer MIPS imaging.  
Over half of our total sample (spectroscopic and photometric redshift
combined) do not have measured redshifts or [OII] emission line 
measurements. As such, we are forced to utilise the 24$\mu$m photometry 
to measure the star formation rate for the bulk of our systems.  As
we only have MIPS imaging in the EGS, we limit our analysis of
star formation rates to galaxies in this field.  Our star
formation measures utilise the [OII] line, when available,
added to the star formation measured from the 24$\mu$m fluxes.
When [OII] is not available, the star formation rate is measured
just using the MIPS fluxes.

To utilise the 24$\mu$m fluxes, we first convert the
24$\mu$m flux into a total IR luminosity L$(8-1000 {\rm \mu m})$ utilising
templates from Dale \& Helou (2002), as parameterised by 
LeFloc'h et al. (2005).    We then 
calculate the obscured star formation rate in these systems through the 
equation,

\begin{equation}
\psi ({\rm IR}) ({\rm M_{*}~yr^{-1}}) = 9.8 \times 10^{-11} \times {\rm L_{IR}}, 
\end{equation}

\noindent which was derived in Bell et al. (2005). Equation
(4) uses a Kroupa/Chabrier
initial mass function, the same as we have used to calculate our
stellar masses.  

As is becoming clear, it is
possible that both the [OII] and the MIPS fluxes could result
from AGN emission.  In the following analysis we have therefore
removed galaxies which have been detected in X-rays, to avoid
the most egregious cases of AGN contamination from our sample.
Out of our sample of 1151 MIPS sources with M$_{*} > 10^{11}$ \solm
at $0.4 < z < 1.4$, we find that 105 or $\sim10$\% are detected
in X-rays. The X-ray properties of our sample are discussed in \S 4.5. 

The star formation rate measured using the [OII] line is 
calculated through the equation,

\begin{equation}
\psi ([{\rm OII}]) ({\rm M_{*}\, yr^{-1}}) = 10^{-11.6-0.4(M_{B}-M_{B \odot})} \times {\rm EW_{[OII]}},
\end{equation}

\noindent where we are forced to use the equivalent width of the line
instead of the flux, as the DEEP2 spectroscopy is not flux calibrated.

Although we do not include sources detected in X-rays in our star formation
analysis, there
is still some chance that the [OII] emission arises from AGN (Yan
et al. 2006).  This is mostly an issue for line emission from
red galaxies, and usually these sources are LINERs.  
We cannot utilise line ratios to test
this idea, as we do not have
 H$\alpha$ or [NII] equivalent widths. We can however
examine the equivalent width ratios of the [OIII] and [OII] lines.
We find that this ratio, [OIII] $\lambda 5007$ / [OII] $\lambda 3727$
is nearly always in the region of star forming galaxies within
our sample. This is consistent with the idea
that bluer galaxies, which dominate our sample at these higher
redshifts, have line ratios consistent with star formation (Yan
et al. 2006).

Perhaps surprisingly, about half of our massive galaxies are detected at
24$\mu$m.  After matching the MIPS and [OII] star formation indicators,
we find that $\sim$40\% of the M$_{*} >$ \mass systems at 
$0.4 < z < 1.4$ are detected to our 24$\mu$m depth.  A total of 
37$\pm5$\% of the systems at M$_{*} >$ \hmass are detected
at 24$\mu$m, with an average star formation rate of 70 \solm yr$^{-1}$.
For systems within the mass range \mass~$<$~M$_{*}$~$<$~\hmass, we
find a higher fraction of MIPS detected systems, with a fraction of
45$\pm$1\%, at an average star formation rate of $\sim 40$ \solm
yr$^{-1}$.

When we compare star formation rates derived from the [OII] line
with those from the 24$\mu$m flux, we find that the 24$\mu$m derived
star formation rates are always much higher. Furthermore, we find little
correlation between the two indicators. This is an indication that [OII]
and IR star formation indicators are measuring two different aspects
of the star formation in a galaxy, or that one, or both, of these methods
is mis-measuring the star formation.  A more detailed discussion of 
different star formation indicators using emission lines and MIPS fluxes
is included in Weiner et al. (2006). 

For galaxies with stellar masses \mass $<$ M$_{*}$ $<$ \hmass,
we find that the fraction of galaxies undergoing star formation
remains roughly similar at all redshifts. Circumstantially, this is
consistent with the fact that the fraction of spirals+peculiars
in this mass cut is roughly constant throughout this redshift
range. Interestingly, the fraction of systems which are undergoing 
star formation is higher than the non-elliptical fraction, showing that
some morphologically classified massive ellipticals must be undergoing star
formation.  

There is a slight decline with redshift
in the fraction of M$_{*} >$ \hmass galaxies with a significant 
24$\mu$m detection.  The fraction declines
from 33\% at $1.2 < z < 1.4$ to 14\% at $0.4 < z < 0.6$, which is
consistent with a drop in the morphological fraction of non-ellipticals. 
This is however certainly a lower limit to the evolution, as the number
of galaxies we can detect at 24$\mu$m declines at higher redshifts.  The
redshift distribution of star formation rate, as measured through
the combined star formation seen in the infrared and [OII]
line is shown in Figure~12, for systems with M$_{*} >$ \mass.

As Figure~12 shows, we are incomplete in measuring
the star formation rate at the highest redshifts of our sample. This
is due to the MIPS imaging becoming incomplete
at the highest redshifts.  When we derive the total star formation
history, we correct for this by assuming that the relative
star formation rate distribution of our sample, that is the relative
number of galaxies at a given star formation rate, is the same at all 
redshifts. This does not assume that the normalisation of the distribution
is the same, simply that the shape of the star formation distribution
function is similar at high and low redshift. This correction
is however small, and only accounts for a 30-40\% increase
in the total star formation rate density.

\begin{figure} 
\includegraphics[width=84mm]{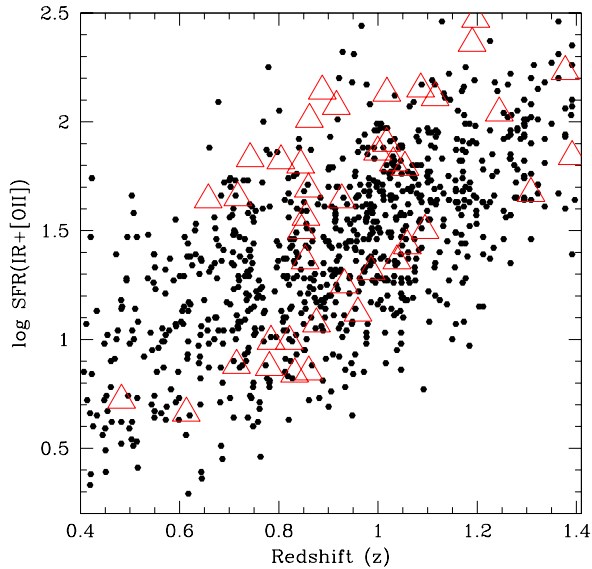}
 \caption{The total star formation rate, found by adding the MIPS infrared
fluxes and [OII] line equivalent widths, as a function of redshift for our
sample.   The large open red triangles show the location of galaxies
with M$_{*} > 10^{11.5}$ \solm, while the solid dots show the
location of galaxies with \mass $<$ M$_{*}$ $<$ \hmass.}
 \label{sample-figure}
\end{figure}

We determine how the star formation rate has evolved
within our sample by examining how the total integrated
star formation density changes as a function of time
and stellar mass (Figure~13).  We find, similar to previous studies
utilising IR star formation indicators (e.g., Le Floc'h
et al. 2005), a decline at lower redshift for our sample,
as seen in the entire field  galaxy population 
(e.g., Lilly et al. 1995) out to $z \sim 1$.  The less massive galaxies 
have a higher average star formation rate per unit time, and have a softer
more gradual decline at lower redshift, compared with the M$_{*} >$ \hmass 
galaxies.  

When we fit these star formation histories up to their
plateau (i.e., at $z \sim 1$) as a power law $\sim (1+z)^{\alpha}$
we find differences between our two mass ranges. For
systems with M$_{*} >$ \hmass, we fit $\alpha = 6\pm2.2$,
and for systems with \mass $<$ M$_{*}$ $<$ \hmass we
fit $\alpha = 4.1\pm0.64$.  The overall decline in the
entire galaxy population's star formation history can be
parameterised as $\alpha = 3-4$ (Hopkins 2004; Le Floc'h et al. 2005). 
It appears that while
the \mass $<$ M$_{*}$ $<$ \hmass galaxies have a similar
decline as the overall field, the highest mass galaxies
show a marginally faster decline.

\subsubsection{Stellar Mass Growth from Star Formation}

We can use our measured star formation rate densities to determine how much
stellar mass is added to our massive galaxy sample through the star formation process.
As we are averaging over the entire sample, we do not need to worry
about the particular time-scales of the star formation induced in
these systems, although the star formation time-scales are likely long (Noeske et al.
2007a,b). In these 
calculations we assume that the star formation rate density remains the same
through a redshift bin. Thus, we can calculate the amount of new mass in
stars created through the star formation process simply by
integrating the star formation rate density throughout the redshift bin of
interest. 
This calculation is independent of whether the star formation
is produced in multiple short bursts, or in stochastic 
star formation occurring through time, as both scenarios
give the same result.

Figure~14 shows the stellar mass evolution, and the amount of mass added to 
our sample due to star formation, as
a function of redshift. This figure is constructed with the mass
change referenced to the stellar
mass density at $z \sim 1.3$, in each bin. The solid and short-dashed lines
show the amount of stellar mass added due to star formation
within the same bins, to galaxies with
 \mass $<$ M$_{*}$ $<$ \hmass and M$_{*} >$ \hmass,
respectively, as a function of redshift.

While the stellar mass increase with time due to star formation
matches within 1-3 sigma the amount of mass at redshifts $z < 1.3$
it is not clear that, when considered together, whether the star
formation by itself can account for changes in the mass
function with time.
We examine this statistically by determining the 
probability that the stellar mass increase due to star formation
can account for the mass changes seen at all redshifts. This
test shows that the probability that star formation, within a bin,
is alone responsible for the stellar mass increase is 0.13\%, at 3 $\sigma$,
for the \mass $<$ M$_{*}$ $<$ \hmass systems, and $0.0025$\% 
(3.5 $\sigma$) for the  M$_{*} >$ \hmass systems. 
Therefore, 
statistically  star formation 
by itself, within a bin, cannot account for the observed
mass changes at the highest redshift bins.  We however do not
consider here the effects of galaxy transfer between
mass bins, although we do so in \S 5.1.  Note that although
statistically the changes in the mass function are not
significant at $> 3 \sigma$, there is a general trend such
that the mass density increases, particularly at $z > 1$. Our
goal here is to determine whether the stellar mass created
from the star formation is consistent with the observed
change.

The stellar mass added from
star formation is therefore unlikely able to account for the changes
in the stellar mass density seen as a function of stellar mass.
This result is however dominated by the large error bars on the mass 
density.  As the only other way to obtain stellar mass growth  is
through galaxies entering bins from lower masses and merging, we 
investigate these processes and derive statistically how much merging
might be occurring from
$z \sim 1.3$ to $z \sim 0.4$ in \S 5.1.
 
\begin{figure}
\includegraphics[width=84mm]{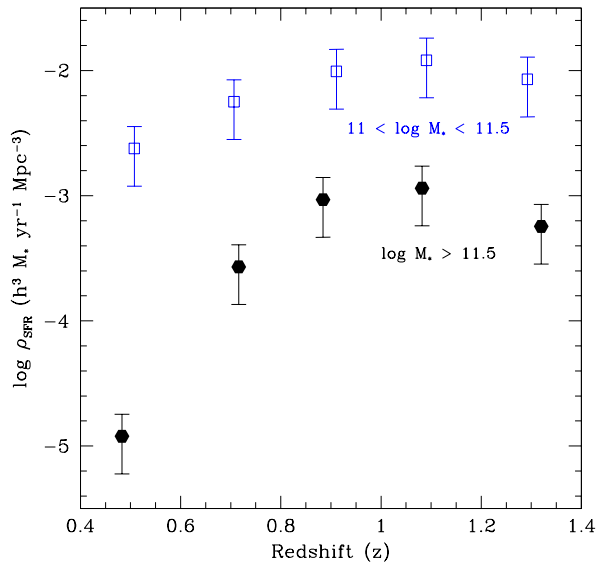}
 \caption{The star formation rate density ($\rho_{\rm SFR}$) as
a function of redshift up to $z \sim 1.4$.  Plotted are the 
measured star
formation rates for both the  \mass $<$ M$_{*}$ $<$ \hmass,
and the M$_{*} >$ \hmass samples.  }
 \label{sample-figure}
\end{figure}

\subsection{Active Galactic Nuclei}
 
To determine the unobscured AGN properties of our massive galaxy
sample we match our galaxies with a preliminary catalog
of X-ray fluxes from Nandra et al. (2007, in preparation). The Nandra et al.
(2007) study is a major Chandra program to cover an area of
0.6 deg$^{2}$ in the EGS with a depth per pointing of 200 ks in
both the hard (3.1 \AA, 3 keV) and soft (12.4 \AA, 1 keV) bands. 
A basic description of this X-ray data is given in Nandra et al.
(2007), Georgakakis et al. (2007) and Davis et al.
(2007), with a full description to appear in Nandra et al. (2007).

We matched our stellar mass selected
catalog to this X-ray catalog, with a valid match occurring for those
within 1-3\arcsec.  There are 123 galaxies in our sample
of M$_{*} >$ \mass that have a matching X-ray detection down to
a limit of 8.2 $\times 10^{-16}$ erg s$^{-1}$ in the hard band,
and 1.1$\times 10^{-16}$ erg s$^{-1}$ in the soft band. 
This accounts for a surprisingly high fraction (5\%) of the entire
M$_{*} >$ \mass sample.  To these limits, there are also
eight galaxies with M$_{*} >$ \hmass with an X-ray detection. 
In both ranges these X-ray sources are found in roughly 5\% of the sample.

The properties of X-ray sources in deep extragalactic near
infrared and optical surveys have been discussed in several
previous papers (e.g., Hornschemeier et al. 2003; Grogin et al. 2005; 
Lehmer et al. 2005; Georgakakis et al. 2006; Nandra
et al. 2006).  These previous studies
have generally found that X-ray sources are in evolved, concentrated
galaxies  up to $z \sim 4$ (Grogin et al. 2005; Lehmer et al. 2005).  This may
imply that X-rays are tracing only one phase of the evolution of
an active galaxy, yet for the purposes of this paper we examine these
X-ray sources with the assumption that they probe a representative
sample of galaxies with active galactic nuclei.

\begin{figure}
\includegraphics[width=84mm]{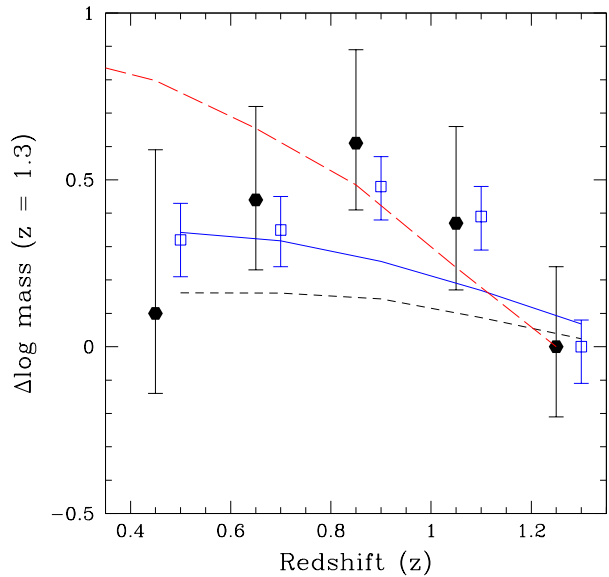}
 \caption{The evolution in the mass density as a function of
redshift and stellar mass.  The solid dots and open boxes
represent the evolution of galaxies with stellar masses  M$_{*} >$ \hmass
and \mass $<$ M$_{*}$ $<$ \hmass, respectively, referenced to
their stellar mass densities at $z \sim 1.3$.  The solid
blue line shows the evolution in the amount of stellar
mass added from the $z \sim 1.3$ bin as a function of redshift from
galaxies with stellar masses of \mass $<$ M$_{*}$ $<$ \hmass.
The short-dashed line shows a similar increase in stellar
mass due to star formation for the M$_{*} >$ \hmass systems.  The 
long-dashed red line shows the relative increase in the stellar
mass for the M$_{*} >$ \hmass systems due to stellar mass brought up
from the \mass $<$ M$_{*}$ $<$ \hmass bin due to star formation.}
 \label{sample-figure}
\end{figure}

Since we are interested in the spectral properties and the luminosities
of the AGN in our sample, we convert our measured X-ray fluxes from the 
Nandra et al. (2007) catalog into X-ray luminosities utilising the formula:

\begin{equation}
{\rm L_{x}} = {\rm flux (2-10 keV)} \times 4\pi D_{L}^{2} (1+z)^{\Gamma-2},
\end{equation}

\noindent where D$_{\rm L}$ is the luminosity distance and $\Gamma$ is
a quantification of the spectral shape.  To obtain a flux which
is independent of intrinsic absorption we calculate our luminosities
using $\Gamma = 1.9$, the intrinsic spectral shape of an AGN, and always
use the hard band flux (2-10 keV), as it is the least affected
by absorption.  

We also investigate the hardness ratios (HR),
$${\rm HR = \frac{(counts_{hard}-counts_{\rm soft})}{(counts_{hard}+counts_{soft})}}$$ of 
our X-ray sources, finding a range of values, with an average 
hardness ratio  of $-0.05\pm$0.58 for the full sample at M$_{*} >$ \mass.  
However, we find that the most massive systems with M$_{*} > $ \hmass have an 
average hardness ratio of -0.50$\pm$0.38, showing that, on average,
the most massive 
galaxies have soft spectra, likely from unobscured AGN. This may
be a further sign of downsizing, where the AGN in the higher mass galaxies
becomes less obscured due to the removal or dissipation 
of the obscuring mechanism earlier than in lower mass galaxies.

Figure~15 shows our basic X-ray analysis results in a graphical format, 
where the stellar masses of our galaxies with an X-ray detection in the EGS
are plotted with their X-ray luminosities, as defined in equation (4).
The morphological types and hardness ratios of these sources are
also labelled.  The X-ray sources are also circled on the CAS
diagrams in Figure~9.  

We generally find no correlation between
X-ray source hardness ratio, X-ray luminosity or host galaxy
stellar mass. However, our range in stellar mass is limited, and
this should not be interpreted as a complete study of AGN, but simply
as a study of the X-ray sources associated with massive galaxies.

There is one interesting trend, which is that the X-ray sources in massive
galaxies tend to be found in peculiars and distorted early-types.  For
the overlapping sample of 29 galaxies with M$_* >$ \mass
that have both X-ray and HST imaging, we find that $\sim 45\pm12$\%
are either peculiar galaxies or distorted ellipticals.  The
remainder are 34$\pm11$\% normal elliptical/S0/compact and 21$\pm8$\%
disks.  This becomes even more significant when we
determine what fraction of each morphological type is detected
in the X-rays. When we do this comparison, we find that
$\sim$15\% of the distorted ellipticals are X-ray detected, while
only 3.5\% of the E/S0/compact galaxies are detected.

This may appear to disagree with previous studies, such
as Grogin et al. (2005) who found that X-ray sources
are largely in low asymmetry and highly concentrated galaxies.  
We find that some of our X-ray sources have a large
concentration index (Figure~9), and some of
the peculiar ellipticals have a modest asymmetry, but
they are not in the merger region at $A > 0.35$.  It
is likely that these earlier studies missed the fact that
a significant fraction of the X-ray sources in the
highest mass galaxies are found in early-type galaxies
with peculiar structures.   Since it is likely that these
structures were produced in some type of a merger several
Gyr earlier, we are potentially witnessing
the remnant AGN activity produced in the last merger event.
These earlier studies of AGN morphology also were not mass
selected, as is our sample.
It is furthermore possible that we are not able to detect
the ongoing peculiar galaxies merging, in X-rays, due to
deep obscuration, although this idea needs to be tested
in more detail in a future study.

\begin{figure}
\includegraphics[width=84mm]{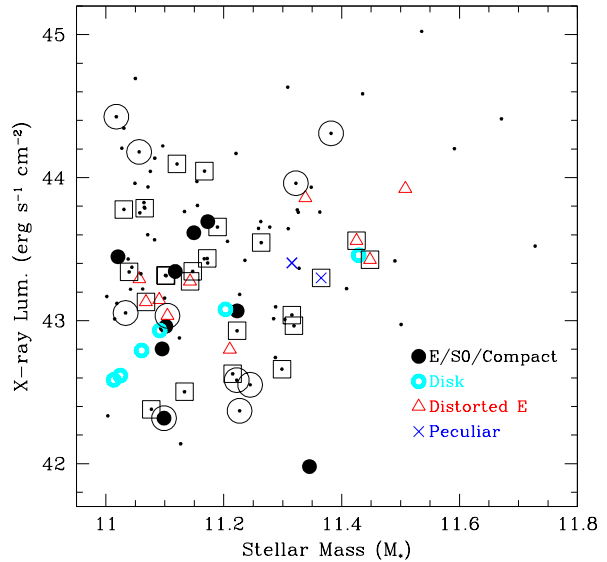}
 \caption{The relationship between the stellar masses of our
X-ray detected
massive galaxy sample and their X-ray luminosities as measured
with Chandra. Sources with corresponding ACS imaging are
labelled using the same scheme in Figure~9.  Sources with
soft hardness ratios $< -0.5$ are circled, while those with
hard ratios $> 0.5$ are squared.}
 \label{sample-figure}
\end{figure}

\section{Discussion}

The last few years have seen a number of studies examine similar
issues as this paper, with many studies concluding that  massive
galaxies, typically those with M$_{*} >$ \mass, are
largely formed by $z \sim 1$.   This is statistically found to be
the case in this paper, to within a factor of 2-3, in terms of measured
mass and number densities, for systems 
with M$_{*} >$ \mass from $z \sim 1$, although there is some evolution from
$z \sim 1.4$.
A scenario where the number and mass densities of M$_{*} >$ \mass galaxies
does not change at $z < 1.4$ can be ruled out at $> 8$ $\sigma$ confidence.

While we, and others, have found that the number and mass densities for
massive galaxies are similar at $z < 1$, this does not necessarily imply
that there is no evolution in this population. As we have discussed,
there can be as much as a factor of three in evolution which we could
not identify due to the uncertainties in the process of measuring stellar
masses, and computing their number densities.  It is likely that other
methods besides densities, are need to conclusively argue whether massive
galaxies are finished forming by $z \sim 1$.  While we can rule out
increases in densities that are a factor of three or greater, we cannot
conclude that our that massive galaxies are not increasing in number and
mass density by a factor of 2 or 3.  The fact that we find a high fraction of
massive galaxies that are forming stars (40\%), and which are non-elliptical
(30\%), at $z \sim 1$ alone suggests that there is some evolution.

One issue which is not clear is how much of the merger and star formation
process, and especially the controversial and hard to find dry mergers, are 
responsible for the addition of mass in massive galaxies at these redshifts.  
We can address this using just our mass functions, and the measured 
star forming histories for galaxies with M$_{*} >$ \mass, and 
\mass~$<$~M$_{*}~<$~\hmass.  While the star formation history matches the
observed increase in the stellar mass to within $< 3~\sigma$ at any
one redshift, the fact
that the star formation history is consistently lower at all redshifts implies
that star formation, within a bin, cannot account for the total increase
in stellar mass.   This shows that part of the mass 
growth in these systems must be accounted for by mergers, or as
we examine below, galaxies with masses lower than the $10^{11.5}$ \solm 
stellar mass
limit evolving into the higher mass bin due to star formation and/or
merging.

We find a higher probability that star formation can account for
the increase in the mass density for systems with \mass $<$ M$_{*} <$\hmass,
with a 0.13\% chance ($\sim$ 3 $\sigma$).  It appears however that
for all galaxies with M$_{*} >$ \mass, additional mass is added to
these redshift bins over that which is seen in the star formation. 
Between the bins M$_{*} >$ \hmass and  \mass~$<$~M$_{*}~<$~\hmass
we can determine the mass transfer due to star formation.
We calculate this in the next section after making a few assumptions 
concerning the distribution of galaxies within these bins, and how the star 
formation seen within these bins is distributed.  

\subsection{Evolution of Number Densities}

In this section we examine ways in which galaxy mass functions can
evolve due to star formation and merging. We consider star formation
within a given mass bin, star
formation at a lower mass bin, and mergers within a lower mass
bin. Note that mergers within a given bin will never increase
the total mass within that bin, unless star formation is
triggered with the merger.
It can however lower the number densities of galaxies within a bin.

Consider the change in the total stellar mass within
a mass range of $\delta M$ at a mass of M$_{0}$. We can
express the evolution of the mass function within  $\delta M$ 
between redshifts $\delta z = z_{2} - z_{1}$ as,

\begin{equation}
\Delta M(M_{0} + \delta M, \delta z) = \Sigma M_{\rm sf} + \Sigma M_{\rm merg}^{\uparrow} + \Sigma M_{\rm sf}^{\uparrow},
\end{equation}

\noindent where $\Delta M$ is the change in the total amount of
mass density within the mass bin between $M_{0}$ and $M_{0} + \delta M$,
and $z_1$ and $z_2$. 
$\Sigma M_{\rm sf}$ is the
amount of mass density within this bin added due to star formation within
this mass bin,
$\Sigma M_{\rm merg}^{\uparrow}$ and $\Sigma M_{\rm sf}^{\uparrow}$
are the amounts of stellar mass density
added to the $M_{0} + \delta M$ bin from mergers, and star formation which were
initially in galaxies at $M_{*} < M_{0}$.  Likewise, these terms should
include mass density which has left the bin due to mergers and star formation.
If we consider the M$_{*} >$ \hmass bin, or any bin which is
defined to be simply greater than a given mass, this additional term is zero.

These terms can be computed in the follow ways. For the star formation
within the bin, where galaxies have an initial stellar mass
between $M_{0} <$ M$_{*} <$ $M_{0} + \delta M$, we can write 
the mass added from star formation as,

\begin{equation}
\Sigma M_{\rm sf} = \Sigma \psi_{\rm sf} \cdot \tau(z_2 - z_1). 
\end{equation}

\noindent Where in equation (8), $\psi_{\rm sf}$ is the star formation rate
density and $\tau$ is the time interval between redshifts
$z_{1}$ and $z_{2}$.  This is the quantity plotted as the short-dashed and
solid lines in Figure~14, and which falls short of accounting
for the possible growth of stellar mass within these massive galaxies (\S 4.4.3).

Because the star formation within these bins cannot account
for the change in the number and mass densities for massive
galaxies (\S 4.4.3), the increase in stellar mass must be brought 
about from galaxies that were previously at lower masses.  This
increase must  be produced by mass entering the
bin through star formation or merging. Since star formation is
in principle easy to directly study and trace, we will examine this
aspect first.

To examine how much stellar mass is added to a bin of lower
stellar mass, we must
consider the star formation rate in galaxies in the lower mass bins. If
the width of a lower stellar mass bin is $\delta M_{\rm lower}$, then the 
amount of
mass added per galaxy ($\delta M_{\rm sf,lower}$) within a redshift and 
mass bin is given by

\begin{equation}
\delta M_{\rm sf,lower} = \frac{\psi_{\rm sf,lower} \cdot \tau(z_2 - z_1)}{\phi_{\rm lower}},
\end{equation}

\noindent where $\phi_{\rm lower}$ is the number density of galaxies in
the lower mass bin.  If we 
assume that galaxies are distributed evenly within their
mass bin, $\delta M_{\rm lower}$, then the fraction of galaxies within this
bin which leave the bin due to star formation is simply 
$\delta M_{\rm sf,lower}/\delta M_{\rm lower}.$ 

 The number density
of galaxies
that move into the higher mass bin due to star formation is then,

\begin{equation}
\delta \phi = \frac{\delta M_{\rm sf,lower}}{\delta M_{\rm lower}}\phi_{\rm lower} = \frac{\psi_{\rm sf,lower} \cdot \tau(z_2 - z_1)}{\delta M_{\rm lower}}
\end{equation}

\noindent Since we assume that the original galaxies are
distributed evenly throughout a bin, then the fraction of
galaxies in the lower mass bin which move into the higher mass bin
is similar to the fraction of mass that moves up. In this case, the
average mass of the galaxies brought up into the higher mass
bin will be $0.5 (M_{\rm top} - \delta M_{\rm sf,lower})$.
Where $M_{\rm top}$ is the upper mass limit of the lower mass
bin from which these galaxies are taken from.  The amount of
mass added to the higher mass bin due to star formation bringing
up galaxies is therefore,

\begin{equation}
\Sigma M_{\rm sf}^{\uparrow}  =  0.5(M_{\rm top} - \delta M_{\rm sf,lower}) \cdot \delta \phi + \Sigma M_{\rm sf,lower} \frac{\delta M_{\rm sf,lower}}{\delta M_{\rm lower}},
\end{equation}

\noindent where $\Sigma M_{\rm sf,lower}$ is the total amount of
star formation occurring in galaxies in the lower mass bin. In this
case, $\Sigma M_{\rm sf,lower} = \psi_{\rm sf,lower} \cdot \tau(z_2 - z_1)$
(cf. eq. 8).

To determine how star formation in 
lower mass bins can affect the mass and number densities in higher mass
bins we take the example of galaxies in the 10$^{11} - 10^{11.5}$ \solm
bin moving into the M$_{*} >$ \hmass bin between $z \sim 1.3$ and $z \sim
1.1$.

In this case 
$\Sigma M_{\rm sf,lower} = \psi_{\rm sf,lower} \cdot \tau(z_2 - z_1)$ = 
$4.7 \times 10^{6}$ h$_{70}^{3}$ \solm Mpc$^{-3}$, and $\delta M_{\rm sf,lower} =
\psi_{\rm sf,lower} \cdot  \tau(z_2 - z_1)/\phi_{\rm lower}$ = 
4.3$\times 10^{10}$ \solm. 
Since $\delta M_{\rm lower} = 10^{11.5}$ \solm $- 10^{11}$ \solm  = 2.2 
$\times 10^{11}$ \solm, 
then the fraction of galaxies in the lower mass bin at $z \sim 1.3$
which are now in the upper mass bin at $z \sim 1.1$ is 
$0.195 \times \phi_{\rm lower}$.  The
change in number density in the upper mass bin, in our example, 
M$_{*} >$ \hmass, is then 4.3$\times 10^{-5}$ Mpc$^{-3}$ dex$^{-1}$.
This is nearly entirely the difference between the number densities for
galaxies at $z \sim 1.3$ and $z \sim 1.1$ for galaxies
with masses M$_{*} >$ \hmass. 

\subsection{Mass Density Evolution and Merging}

We calculate in the previous section that in the highest redshift bin there
is no need for additional galaxies due to merging using the 
observed change in number densities. However,
a better test of the merger scenario is to examine how the total mass
densities evolve.  The reason
is that when mergers occur within a given mass
bin the number density will decrease, but the mass density remains
conserved, if no star formation is triggered. Because star formation 
and perhaps merging is occurring
within lower mass galaxies, those systems will enter the higher
mass bin and increase the number of galaxies, creating a 
static number number density, masking real evolution. However,
in this case the integrated mass density must increase within
our observed bin.

We continue our examination of mass transfer between the M$_{*} >$ \hmass
and \mass $<$ M$_{*} <$ \hmass mass bins.   We calculate the amount of
mass added to the M$_{*} >$ \hmass bin by subtracting the amount of mass
added from star formation from the total mass density change.
 The amount of stellar mass added 
to the M$_{*} >$ \hmass
bin from star formation in the \mass~$<$~M$_{*}~<$~\hmass bin
is given by eq. (11). In this case, $\Sigma M_{\rm sf}^{\uparrow}$ 
= 3.9$\times 10^{6}$ h$_{70}^{3}$ \solm Mpc$^{-3}$. The total
integrated stellar mass added to galaxies at M$_{*} >$ \hmass
due to galaxies at \mass~$<$~M$_{*}~<$~\hmass moving into
the M$_{*} >$ \hmass limit from star formation is shown in Figure~14 as
the long-dashed red line.  

This additional mass accounts
for roughly half of the mass increase in the  
M$_{*} >$ \hmass bin up to $z \sim 0.8$,
with the remainder of the mass possibly originating from
merging in the upper part of the \mass~$<$~M$_{*}~<$~\hmass bin. 
We however cannot rule out that the scenario that mergers are not
needed to account for the change in mass densities as the star
formation from the lower mass bin is only lower than the change
in mass densities by 1$\sigma$.
By subtracting the star formation within the M$_{*} >$ \hmass
bin, and the mass added from star formation in the lower mass bin, we
calculate $\Sigma M_{\rm merg}^{\uparrow}$ = 
3.3 $\times 10^{6}$ h$_{70}^{3}$ \solm Mpc$^{-3}$ using eq. (7).  

To calculate the number of possible mergers, and the merger
fraction within the \mass~$<$~M$_{*}~<$~\hmass bin,
we first must recognise which types of mergers, and
of what masses, can enter the M$_{*} >$ \hmass bin.
First, to simplify things, we assume that galaxies
within a given mass range and redshift range undergo
at maximum one merger between $z_{1}$ and $z_{2}$. This
is a very reasonable assumption within our redshift bins
given that even the fastest
mergers take around 0.5 Gyr to complete. We furthermore assume that all 
mergers within the lower mass bin are major mergers of
equal mass. With these assumptions, it is clear
that only galaxies with M$_{*} >$ $M_{\rm top}$/2
will have a mass  M$_{*} >$ $M_{\rm top}$ after
merging with a galaxy of similar mass.  Therefore,
only galaxies merging with masses between M$_{*} = M_{\rm top}$/2, and
 M$_{*} = M_{\rm top}$ will enter the upper mass bin.  If we
assume that galaxies are evenly distributed between these
two masses, then the average mass of a merging galaxy
is $0.5 (M_{\rm top}/2 + M_{\rm top})$, or
3/4$M_{\rm top}$.  The total number density of galaxies 
merging\footnote{Note that the total number of galaxies
merging is not the same as the total number of galaxy mergers,
which is a factor of two lower.}, ($n_{\rm mg}$), is therefore equal to
the stellar mass density of galaxies which have merged divided
by the average mass of the merging galaxies,

\begin{equation}
n_{\rm mg} = \frac{4\Sigma M_{\rm merg}^{\uparrow}}{3M_{\rm top}}.
\end{equation}

\noindent  We can compute the galaxy
merger fraction (Conselice 2006), $f_{\rm mg}$, as the total number of
galaxies undergoing a merger divided by the total number
of galaxies within a given bin.  This fraction can then be
expressed as,

\begin{equation}
f_{\rm mg} \approx \frac{4 \Sigma M_{\rm merg}^{\uparrow}}{3 (\phi_{\rm lower})(M_{\rm top})}.
\end{equation}

\noindent Plugging in numbers for our \mass $<$ M$_{*} <$ \hmass and
M$_{*} >$ \hmass mass bins, we find that 12\% of the galaxies with
\mass~$<$~M$_{*}~<$~\hmass merge between $z \sim 1.2 - 1.4$, and thus
enter the higher mass bin.  We find that at $z \sim 0.8 - 1.0$ this
merger fraction drops to 8\%.  This is consistent with the CAS
results presented in \S 4.3.5, and previous published results
(Conselice et al. 2003; Lin et al. 2004; Bundy
et al. 2004; Lotz et al. 2006; Bridge et al. 2007). This suggests
that although number and mass densities do not statistically differ
from each other, by examining the mass function in narrow redshift
bins as well as the star formation rate and morphologies we
conclude that there is additional mass added to these
systems, by as much as a factor of two increase since $z \sim
1.4$, which is largely due to merging.  Previously, we 
could not rule out no change in number densities, based only
on counting galaxies.  The evolution is however clearly
seen in the structures of these galaxies.

\subsection{Comparison to models}

We also compare our stellar mass and number densities  results to models 
from e.g., De Lucia et al. (2006) based on the Millennium
Simulation (Lemson et al. 2006) as another approach towards
understanding how well we can reproduce with simulations the evolution of 
these systems. The Millennium Run is a $\Lambda$CDM 
simulation of the universe using 10$^{10}$ particles and follows the dark 
matter and stellar assembly of galaxies in a 500$h^{-1}$ Mpc cube. It is
currently one of the most advanced simulations that are available for
comparison to data, and have been used previously to trace
the evolution of massive galaxies (De Lucia et al. 2006).  

While the $\lambda$CDM model is very successful in explaining the large
scale structure of the universe, there are inconsistencies
when comparisons are done on the scale of galaxies.  By testing this model in
other ways, such as through the assembly of the most massive galaxies,
we can perhaps determine the origin of some of the missing physics, or
initial conditions used in these simulations.

The stellar mass calculation in the Millennium simulation uses the same IMF 
as our 
stellar mass measurements, and are based purely on the amount of stars 
produced in galaxies as a function of time.  The Chabrier IMF enters into 
these models through the computation of the fraction of gas in star 
formation which is returned to the cold gas, which in these models is 40\%.
There is also a prescription for which stars explode as supernova,
and when, which further reduces the stellar mass, but only for
the stars at the highest masses.

When we compare our stellar mass number and mass densities to the
Millennium model, we find that they generally underestimate
the number and mass densities of the most massive galaxies at
$z < 2$ (Figure~16). The agreement for both stellar mass number
and mass densities at $z \sim 0$ is however fairly good within the 
uncertainties.  The best agreement is between the number densities
of systems with \mass $<$ M$_{*}$ $<$ \hmass. However, as can
be seen in Figure~16, this agreement breaks down when comparing
the mass densities of these systems to the model.  As there is generally a much
worse agreement for the M$_{*} >$ \hmass systems in terms of mass
densities than number densities, we can interpret these disagreements
in the sense that the models do not produce enough massive galaxies early
enough. This extends throughout the M$_{*} >$ \mass range.
Clearly, the models eventually produce these massive systems, as there is
agreement at $z \sim 0$, but they are formed in the simulation
at much later times, typically at $z < 1$ through  mergers 
(De Lucia et al. 2006).  

This disagreement suggests that the bulk formation of massive galaxy formation
occurs at $z > 2$, especially for the  massive systems. This implies
both an early star formation history, as well as an early mass assembly
history.  When comparing with observations of the merger history, it is
clear that massive galaxies are formed through mergers at $z > 2$
(Conselice 2006).  It is likely that these galaxies are much more
biased than what the Millennium simulation assumes, and thus they
must have an early merger and star formation history.   

When
examining the galaxy merger history (Conselice et al. 2003; Conselice
2006), it is clear that massive galaxies do not merge
much at $z < 2$, but have a high merger fraction ($\sim 50$\%) 
at higher redshifts, $z > 2$.  The Millennium model produces 
most of the mass in these massive galaxies by $z \sim 2$, but they are 
not yet assembled in massive galaxies in the model until
relatively late (e.g., De Lucia et al. 2006).  It appears that the 
actual merger
rate for massive galaxies is much higher in the past, than what
is predicted, and likewise smaller at $z < 2$ than the models.

\begin{figure*}
 \vbox to 150mm{
\includegraphics[angle=0, width=154mm]{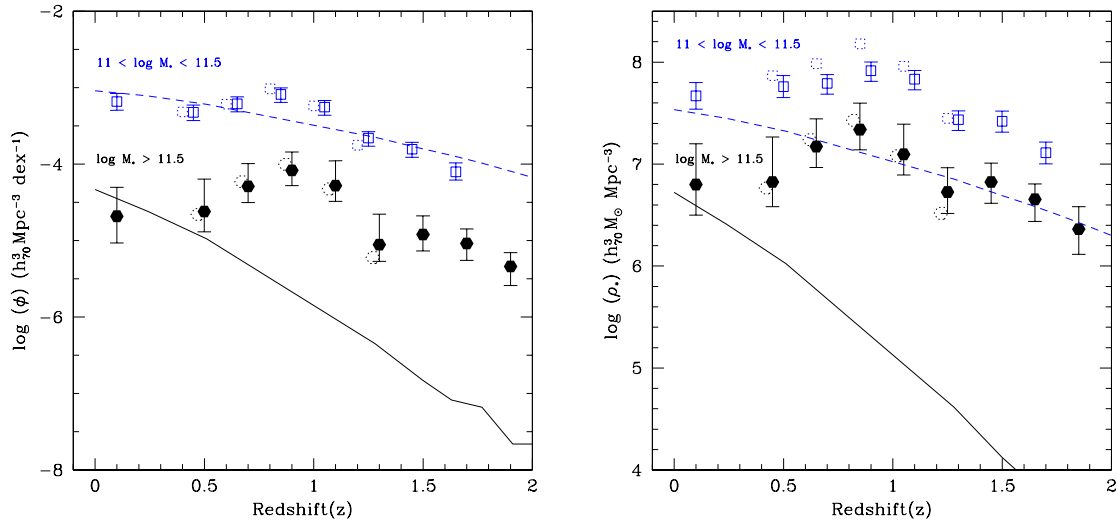}
 \caption{A comparison between our data and the models from
the Millennium simulation (e.g., De Lucia et al. 2006). The data
shown are the same for the massive galaxy sample plotted in
Figure~4.  The dashed line shows the predicted evolution in the
number and stellar mass densities for \mass $<$ M$_{*} <$ \hmass systems,
while the solid line shows the same predicts for galaxies with stellar
masses M$_{*} >$ \hmass.  As can be seen, for the most part the simulated
massive galaxies do not assembly quickly enough to match the observations. }
} \vspace{-5cm} \label{sample-figure}
\end{figure*}

\section{Summary}
  
We utilise wide and deep near infrared imaging from the Palomar
telescope combined with DEEP2 spectroscopy to select and study the
properties and evolution of M$_{*} >$ \mass galaxies found at
$0.4 < z < 1.4$.  Our total sample consists of 4571 galaxies 
with M$_{*} >$ \mass, and 225 galaxies with M$_{*} >$ \hmass.   
We investigate with DEEP2 spectroscopy, Hubble Space Telescope (HST) 
imaging, Chandra imaging, and Spitzer MIPS imaging: the 
X-ray, morphological and 
star forming properties of our sample.  Our major findings are: \\

\noindent 1. The stellar mass and  number densities of M$_{*} >$ \mass
galaxies does not change significantly at $z < 1$.  We however cannot
rule out factors of 2-3 in number and mass density evolution for
these systems, based solely on densities, due to uncertainties
in these measurements.  Systems with
\mass $<$ M$_{*} <$ \hmass however show a significant increase
between $z \sim 1 - 1.5$.
 The increase in stellar mass density for systems
with M$_{*} >$ \hmass is however marginally insignificant over
our entire redshift range up to $z \sim 2$.
\\

\noindent 2. There is a diversity in the morphological properties of
galaxies selected solely by stellar mass.  We find that most galaxies 
selected with M$_{*} > 10^{11.5}$ \solm
are classifiable as early-types at all redshifts, with only
a small fraction classified as peculiars and spirals at $z \sim 1.3$. Systems
with \mass $<$ M$_{*} <$ \hmass have a lower fraction of galaxies
consistent with being early-types, with a nearly constant fraction
with redshift of 70\%.  The remaining systems tend to be
classified as peculiars/mergers at high redshift, and spirals
at lower redshifts.  We further find that a significant number of
the early-type galaxies contain a slight morphological disturbance,
which is furthermore seen in the quantitative CAS parameters.\\

\noindent 3. We find that a significant fraction ($\sim 40$\%) of massive
galaxies at $0.4 < z < 1.4$ are undergoing star formation.  This
is demonstrated through the blue colours of massive galaxies,
and star formation rates as derived through Spitzer MIPS imaging
of these systems.  The fraction of galaxies with  M$_{*} > 10^{11.5}$ \solm
which are on the red-sequence is only 60\% at $z \sim 1.3$, but increases
to $\sim$ 100\% by $z \sim 0.4$. Likewise, the fraction of lower mass systems,
with \mass $<$ M$_{*} <$ \hmass, which are on the red-sequence increases
with time, although at every redshift there is a higher fraction of
systems with  M$_{*} > 10^{11.5}$ on the red-sequence than at lower
masses.\\
 
\noindent 4. We investigate the star formation rate density for
our massive galaxy sample, finding a steep decline in the
star formation rate density for systems with  M$_{*} >$ \hmass,
and a more shallower decline for systems with \mass $<$ M$_{*} <$ \hmass.
The rate of decline in star formation for M$_{*} >$ \hmass systems 
can be parameterised
as $(1+z)^{6\pm2.2}$, as opposed to $(1+z)^{4\pm0.6}$ for 
systems with masses \mass $<$ M$_{*} <$ \hmass.\\
									
\noindent 5. We match our catalog of massive galaxies to a catalog
of X-ray sources discovered with Chandra imaging in our largest
field, the Extended Groth Strip. We find that a significant
fraction (5\%) of the most massive galaxies are X-ray emitting
AGN. We investigate in detail the properties of these AGN
and find that the most massive galaxies, with M$_{*} >$ \hmass,
have soft hardness ratios, while the galaxies with \mass $<$M$_{*} <$ \hmass
have a mixture of hard and soft ratios.  Furthermore, we find that
nearly half of all the massive galaxies which have AGN are either
distorted ellipticals or peculiars. \\

\noindent 6. We investigate how much stellar mass is added to
galaxies due to star formation from $z \sim 1.4$ to $z \sim 0.4$,
and compare this to the observed changes in the stellar mass density. We find
to $> 3~\sigma$ confidence that the star formation seen in individual
massive galaxy bins at $z < 1.4$ cannot account for the changes in stellar
mass seen in galaxies with M$_{*} >$ \mass.  We however find that the
amount of mass which transfers between bins due to
star formation, bringing galaxies up into higher mass bins, can 
within 1 $\sigma$
account for this increase. The difference suggests that up to a single
major merger is occurring for M$_{*} >$ \mass galaxies 
at $0.4 < z < 1.4$.  This is also found to be the case through an analysis
of CAS parameters, giving $0.9^{+0.7}_{-0.5}$ major mergers at $0.4 < z < 1.4$.

Our major conclusion in this paper is that the stellar mass
assembly of massive
galaxies is, within a factor of two, complete by $z \sim 1$. There is however
still significant  morphological and colour
evolution at $z < 1$ for massive galaxies.  We  also find
significant evolution in the number and mass densities for galaxies
with M$_{*} >$ \mass at $z > 1$.
Furthermore, we have shown that the study of `early-types', defined
through colour, morphology or mass, at
high redshifts must be carefully done, and results of studies
will vary significantly, depending on selection.  It is
clear, particularly at high redshift, that red galaxies are
not the equivalent of massive galaxies, or elliptical galaxies,
and each of these populations must be studied individually.  

The Palomar and DEEP2 surveys would not have been completed without the active
help of the staff at the Palomar and Keck observatories.   We thank
Niv Drory, Karl Glazebrook, Gabriella De Lucia, Gerard Lemson \& the 
Virgo Consortium for providing data in an electronic formation, and 
Sandy Faber and the anonymous referee for their comments on this work.  
We thank Stephen Bevan for
computing stellar masses using both the Bruzual \& Charlot (2003) and
(2007) models, and we thank G. Bruzual and S Charlot for providing us
with their 2007 models before publication.
Funding to support this effort came from a National Science Foundation
Astronomy \& Astrophysics Fellowship, grants from the UK Particle
Physics and Astronomy Research Council (PPARC),
Support for the ACS imaging of the EGS in GO program 10134 was 
provided by NASA through NASA grant HST-G0-10134.13-A from the Space Telescope Science Institute, which is operated by the Association of Universities for Research in Astronomy, Inc., under NASA contract NAS 5-26555.  This work is based, in part, on observations made with the Spitzer Space Telescope, which is operated by the Jet Propulsion Laboratory, California Institute of Technology under a contract with NASA. Support for this work was provided by NASA through contract 1255094 issued by JPL/Caltech.

The authors wish to recognise and acknowledge the very significant cultural role and reverence that the summit of Mauna Kea has always had within the indigenous Hawaiian community. We are most fortunate to have the opportunity to conduct observations from this mountain.
 ALC/JAN is supported by NASA through Hubble Fellowship grant HF-01182.01-A/HF-011065.01-A.  Support for this work was also provided
by NASA to CJP through the Spitzer Space Telescope Fellowship Program,
through a contract issued by the Jet Propulsion Laboratory, California
Institute of Technology under a contract with NASA.

\appendix

\section[]{CAS vs. Visual Morphologies}

In this appendix we describe in more detail the degree of agreement between the
CAS and visual estimates of morphology.  We are particularly interested
in describing why some systems classified as a particular type by eye
fall into a different region of CAS space than what their apparent
morphology would indicate.  A more limited discussion of this issue has
been presented in Conselice (1997), Bershady et al. (2000), Mobasher
et al. (2004) and Conselice et al. (2005a).

Before we discuss why there are differences between visual estimates
of morphological type and the quantitative approach, it is important
to describe the process of morphological classification by eye.  This
process was carried out by one of us (CJC) by examining a galaxy's 
structure on a computer screen. This was done through IRAF and 
the viewing tool DS9, although similar results can be had through
any approach that lets the viewer change the contrast.  Classification
was done into the types described in \S 4.3 using the traditional approach
of examining the entire galaxy at once, and judging which of the
categories it belongs to.  When this process is done we do not
take into account the smaller, more subtle, details of a galaxy's structure. For
example, although most of our sample of $>$ \mass galaxies consists of
ellipticals, about a third have some kind of
morphological peculiarity.  Typically, this was either a diffuse outer
envelope of material, or a lopsided centre with respect to the rest 
of the galaxy.  This is clearly seen the CAS space for these systems -
the peculiar ellipticals have higher asymmetry and clumpiness values
than ellipticals, compacts and S0s.

However, there are cases in which a galaxy has been classified as one type
but appears in a region of the CAS space where it ordinarily should not be.
Examples of this include ellipticals with high asymmetries, or low
concentrations, and peculiar galaxies with low asymmetries.  While on
average these galaxy populations are found in the location expected within
the CAS space, there are some obvious exceptions.  After creating
the first version of Figure~9 we went back and examined by eye the
$\sim 5$\% of galaxies whose visual morphology differed greatly
from their measured
CAS values.  About half of the time it turned out that the galaxy
was misclassified by eye or recorded incorrectly.  This always accounted
for the most obvious cases where the CAS values and eye estimates of morphology
differed the most.

There are however systems that still differ in terms of their
visual estimates of morphology and their measured CAS values.  It
turns out that the CAS parameters are most successful at distinguishing
disk galaxies, ellipticals and mergers from each other (Cassatta et al.
2005).  We therefore focus on these three populations. First, perhaps
the most obvious disagreement is the high-A ellipticals and
the low-A peculiars.  The ellipticals classified by eye that contain
a high asymmetry were nearly always systems that had some morphological
peculiarity.     In other cases, it was determined that the high-A
elliptical/S0 systems were nearly edge-on, or otherwise peculiar, S0s that
often contained a prominent dust lane.  Other examples were ellipticals
that contained neighbouring galaxies that created a higher-A signal. 

The peculiars that are at $A < 0.3$ show a diversity of visual
morphologies. Often these systems were compact and concentrated with some
diffuse material, or had blob like features nearby.  In many cases these
systems appear to be ellipticals in assembly, and could have easily been
classified as peculiar ellipticals.  Other examples of lower-A peculiars
are systems imaged with a low-S/N.   As has been argued in Conselice
et al. (2000) and Conselice (2003) a lower S/N will make it more difficult
to measure quantitative indices for these systems.  Several systems
resembled spiral galaxies, or spirals in assembly, with what appears
to be arms in development.  It thus appears that quantitative measures
of structures are revealing information that eye-ball estimates are
missing, namely that a gross morphology can be present while the galaxy
structure still retains
some signatures of recent formation. This creates significant
differences when comparing galaxies of the same type, as
selected by eye-ball estimates, at low and high redshifts.

\label{lastpage}

\end{document}